\documentclass[a4paper]{article}

\usepackage{epsfig}
\usepackage{amssymb,amsmath,amsfonts}
\usepackage{multirow}
\usepackage{rotating}
\usepackage{graphicx}
\usepackage{lscape}
\usepackage{hyperref}
\usepackage{ulem}
\usepackage{xcolor}

\usepackage{units} 

\usepackage{cite} 

\usepackage{authblk} 

\usepackage{geometry}
\begin{document}
\title{\bf
Vectorlike lepton imprints at lepton $g-2$ measurements and $e^+e^-$ colliders	\\
}
\author[1]{Sang Quang Dinh%
}
\affil[1]{VNU University of Science, Vietnam National University - Hanoi,
	334 Nguyen Trai Road, Hanoi, Vietnam
}
\author[2]{Hieu Minh Tran%
	\thanks{hieu.tranminh@hust.edu.vn}}
\affil[2]{Hanoi University of Science and Technology, 
	1 Dai Co Viet Road, Hanoi, Vietnam
}
\maketitle
\begin{abstract} 

A fermion can be chiral or vectorlike with respect to a given symmetry, depending on its coupling to the corresponding gauge boson.
Vectorlike fermions have a distinct property that their left-handed and right-handed components behave in the same way under the gauge symmetry.
In this paper, we investigate an extension of the standard model with an $SU(2)$ doublet of vectorlike leptons and two complex scalars.
The new physics effects on the lepton anomalous magnetic moment, as well as the electron and muon pair production processes at $e^+e^-$ colliders are analyzed.
Taking into account the updated measurement results of the electron and muon $g-2$, the LEP and the LHC data, the viable parameter space of the model is identified.
We also examine the prospect of testing the model using $\mu^+\mu^-$ signals from electron-position annihilation at the Future Circular Collider (FCC-ee).
The analysis shows that the FCC-ee will be able to exclude a significant part of the parameter space, pinpointing exiguous viable regions to be tested in the future due to its high precision.

\end{abstract}
\setcounter{tocdepth}{2} 
\newpage
%
%
%
\section{Introduction}


Fermions can be classified into two categories, chiral and vectorlike, depending on their couplings with gauge bosons.
Vectorlike fermions are beautiful in the sense that their left-handed and right-handed components transforms in the same way under a given symmetry.
They possess a special property that their mass terms can be introduced directly in the Lagrangian without violating any symmetry.
Thus, the Higgs mechanism is not necessary to generate such terms.
Moreover, a model with vectorlike fermions is safe from the triangle anomalies \cite{Georgi:1972bb}.
Charged fermions in the standard model (SM) are chiral with respect to the electroweak gauge symmetry $SU(2)_L \times U(1)_Y$ while they are vectorlike under the color symmetry $SU(3)_C$ and the remnant electromagnetic symmetry $U(1)_{EM}$.
Although, the SM setup ensures that the model  is anomaly free and results in rather good agreement with experiments, it still raises a few open questions: 
\textit{(i)} Why is there such a difference in the chiral/vectorlike behaviors of the SM fermions with respect to different symmetries?
\textit{(ii)} Is it deeply connected to the fact that the electroweak symmetry is spontaneously broken and the color symmetry is preserved?
\textit{(iii)} Is there any other fermion being vectorlike with respect to the whole gauge symmetry of the SM?
Answers to these puzzles are indeed interrelated.
Regarding the second question, if it is the case, fermions must be chiral in a grand unification theory since the corresponding gauge symmetry is broken at low energy scales.
Regarding the third question, if there exists vectorlike fermions with respect to the whole SM gauge symmetry, 
it immediately implies an answer to the second question:
the vectorlike (chiral) property of SM fermions is not strictly related to the preserved (broken) gauge symmetry.

Following the latter direction, many models were built using vectorlike fermions since they can solve certain problems
\cite{DeRujula:1975smg, Fritzsch:1975sr, delAguila:1989rq, 
Cheng:1999bg, delAguila:2000rc,
Arkani-Hamed:2002ikv,
Han:2003gf, Cheng:2005as, 
Kang:2007ib, 
Cacciapaglia:2011fx, Aguilar-Saavedra:2013qpa, Ellis:2014dza, Angelescu:2015uiz, 
Arhrib:2016rlj, Barducci:2017xtw, Cacciapaglia:2018qep, 
Arhrib:2018pdi, Song:2019aav, 
Bhattacharya:2021ltd, CarcamoHernandez:2023wzf, Branco:2022fmj, CarcamoHernandez:2021yev}.
It was shown that the vectorlike fermion contributions to the renormalization group evolution of the gauge couplings can lead to precise unification of these couplings at high energy scales
\cite{Emmanuel-Costa:2005qsv, 
Barger:2006fm,
Dermisek:2012ke, Dorsner:2014wva,
Bhattacherjee:2017cxh, Kowalska:2019qxm, Olivas:2021nft}.
Vectorlike fermions also affect the running of the Higgs quartic coupling, addressing the vacuum stability of the Higgs potential
\cite{Blum:2015rpa, Gopalakrishna:2018uxn, Arsenault:2022xty, Hiller:2022rla, 
Hiller:2023bdb,
Adhikary:2024esf, Cingiloglu:2024vdh}.
The effects of vectorlike quarks in flavor physics, such as the $b\rightarrow s$ transitions 
\cite{AristizabalSierra:2015vqb, Dinh:2020inx, Belanger:2015nma, Belanger:2016ywb, 
Kawamura:2019rth, Cherchiglia:2021vhe}
and the unitarity of the Cabibbo-Kobayashi-Maskawa matrix
\cite{Cheung:2020vqm, Crivellin:2020ebi, Dinh:2023ezl, Alves:2023ufm, Belfatto:2021jhf, Branco:2021vhs, Balaji:2021lpr, Crivellin:2021bkd}, were investigated.
While the Large Hadron Collider (LHC) imposes a strong constraint on vectorlike quarks
\cite{ATLAS:2022ozf, ATLAS:2022hnn, ATLAS:2022tla, ATLAS:2023pja, ATLAS:2023bfh, CMS:2022yxp, CMS:2022tdo, CMS:2022fck, CMS:2023agg},
the constraint on vectorlike leptons is milder \cite{CMS:2022cpe, ATLAS:2023sbu}.
However, these constraints are usually not universal. 
They are only applied to specific classes of models with certain simplified assumptions.

The vectorlike leptons contributions to lepton anomalous magnetic moments has been investigated intensively for more than a decade
\cite{Kannike:2011ng, Dermisek:2013gta, Ishiwata:2013gma, Falkowski:2013jya,
Queiroz:2014zfa, Freitas:2014pua, Aboubrahim:2016xuz, Kowalska:2017iqv, Poh:2017tfo, Choudhury:2017fuu, Choudhury:2017acn, Raby:2017igl, Calibbi:2018rzv, Crivellin:2018qmi, Arnan:2019uhr, Kawamura:2020qxo, DeJesus:2020yqx, Endo:2020tkb, Huang:2020ris, 
Jana:2020joi,
Frank:2020smf, 
Chun:2020uzw, Chakrabarty:2020jro,
Dermisek:2021ajd,
Ferreira:2021gke, Athron:2021iuf, Hernandez:2021iss, Arkani-Hamed:2021xlp, Saez:2021qta, Bonilla:2021ize, Hernandez:2021xet, Ko:2021lpx, Cherchiglia:2021syq, 
Lee:2022nqz, Kawamura:2022uft, Nagao:2022oin, Brune:2022rlo, Dermisek:2022aec,
Raju:2022zlv, Wojcik:2022woa, Hamaguchi:2022byw, Dermisek:2023nhe, CarcamoHernandez:2023wzf, Lourenco:2023sde, Duy:2025jbw, Mohling:2024qvk, Athron:2025ets}.
%
%
%
%
%
Recently, the Muon $g-2$ Collaboration at Fermilab announced the final measurement result of the muon anomalous magnetic moment,
$a_\mu = \frac{1}{2}(g_\mu-2)$,
with the unprecedented precision of 127 ppb
\cite{Muong-2:2025xyk}.
Combined with previous result of the Brookhaven National Laboratory
\cite{Muong-2:2006rrc},
the new world average value for the muon $g-2$ is given by
\cite{Muong-2:2025xyk}
\begin{align}
	a_\mu^\text{exp} = 
	(116592071.5 \pm 14.5) \times 10^{-11},
\label{amuEXP}
\end{align}
corresponding to the overall precision of 124ppb.
The experimental uncertainty is expected to be improved further in the future with the projected measurement of the muon $g-2$ at J-PARC using a different method
\cite{Muong-2:2015xgu, Mibe:2010zz}.
On the theoretical side,
thanks to the efforts of the lattice QCD calculation \cite{Borsanyi:2020mff, Boccaletti:2024guq,
Ce:2022kxy,
ExtendedTwistedMass:2022jpw, 
RBC:2023pvn, 
Kuberski:2024bcj, 
Spiegel:2024dec, 
RBC:2024fic,
Djukanovic:2024cmq,
ExtendedTwistedMass:2024nyi,
MILC:2024ryz,
FermilabLatticeHPQCD:2024ppc
}
of the hadronic vacuum polarization contributions to the muon $g-2$, the SM prediction of $a_\mu$ 
\cite{Aliberti:2025beg, 
Ludtke:2024ase, 
Hoferichter:2025yih, 
Bijnens:2021jqo,
Danilkin:2021icn,
Stamen:2022uqh,
Estrada:2024cfy,
Deineka:2024mzt,
Eichmann:2024glq,
Hoferichter:2024bae,
Holz:2024diw,
Chao:2021tvp,
Chao:2022xzg,
Blum:2023vlm, Fodor:2024jyn, 
Aoyama:2012wk,
Aoyama:2017uqe,
Czarnecki:2002nt,
Gnendiger:2013pva,
Gerardin:2019vio,
Blum:2019ugy
}
becomes closer to the experimental value 
\begin{align}
a_\mu^\text{SM}	=
	(116592033 \pm 62)\times 10^{-11}.
\label{amuSM}
\end{align}
Hence, the long standing discrepancy between the SM prediction and the experimental measurement of the muon anomalous magnetic moment has been resolved.
Although the deviation is smaller, there is still a gap for new physics to play its role, especially when the theoretical uncertainty is reduced in the future \cite{Athron:2025ets}.
%
%
%
Meanwhile, the current most precise measurement of the electron anomalous magnetic moment,
$a_e = \frac{1}{2}(g_e - 2)$, was reported in Ref. 
\cite{Fan:2022eto},
where the value of this quantity is
\begin{align}
a_e^\text{exp} =
	(1159652180.59 \pm 0.13)\times 10^{-12}	.
\label{aeEXP}
\end{align}
The SM prediction of the electron $g-2$
was calculated with five-loop QED contribution
\cite{Aoyama:2012wj, 
	Aoyama:2019ryr, 
	Laporta:2017okg,
	Volkov:2019phy,
	Volkov:2024yzc,
	Aoyama:2024aly
}
and hadronic loop corrections
\cite{Hoferichter:2025fea, DiLuzio:2024sps,
Jegerlehner:2009ry,
Prades:2009tw}.
With the fine structure constant measured most accurately using rubidium atom recoil
\cite{Morel:2020dww}, the SM prediction of the electron anomalous magnetic moment reads
\begin{align}
a_e^\text{SM} (\text{Rb}) =
 (1159652180.252 \pm 0.095) \times 10^{-12},
\label{aeSM}
\end{align}
that is about 2$\sigma$ away from the above experimental value.%
\footnote{Notice that there is an unsolved mystery in the large discrepancy between the fine-structure-constant measurement utilizing rubidium recoil \cite{Morel:2020dww} 
and the one using cesium recoil \cite{Parker:2018vye}.
With the latter experimental result, the SM prediction of the electron $g-2$ is about 3.9$\sigma$ deviated from the experimental value, namely
\begin{align}
	a_e^\text{SM} (\text{Cs}) = 
	(1159652181.61 \pm 0.23) \times 10^{-12}.
\end{align}
}
Together with the muon $g-2$, the anomalous magnetic moment of electron play an important role in constraining new physics that couples to the lepton sector.

To explain the above deviation, 
a charged vectorlike lepton singlet coupled to a lepton doublet and the Higgs of the SM is introduced in the minimal vectorlike lepton extension of the SM \cite{
Queiroz:2014zfa, Kowalska:2017iqv, DeJesus:2020yqx, Cherchiglia:2021syq}.
The contribution to $g-2$ can be enhanced further due to the chirality flipping effect in a more complicated model with both a singlet and a doublet of vectorlike leptons 
\cite{Kowalska:2017iqv, Poh:2017tfo, Crivellin:2018qmi, Arnan:2019uhr, Endo:2020tkb, Chakrabarty:2020jro, Arkani-Hamed:2021xlp, Kawamura:2022uft, Hamaguchi:2022byw, Dermisek:2023nhe, Mohling:2024qvk}.
However, the Yukawa interaction between vectorlike and SM leptons implies a mass mixing among these particles, which in turn leads to non-trivial flavor changing neutral currents (FCNCs) at the tree level in the lepton sector.
To avoid such dangerous FCNCs, instead of using the SM Higgs, one can introduce a new scalar that does not develop a non-zero vacuum expectation value.
%
%
In an alternative approach, only one vectorlike lepton doublet is introduced beside the SM fermions.
%
While the new physics contribution to lepton $g-2$ has similar size to that in the minimal model,
this next-to-minimal setup has richer phenomenology since the new leptons interact with both $U(1)_Y$ and $SU(2)_L$ gauge bosons.
The new physics contributions of this model to the scattering and pair creation processes at $e^+e^-$ colliders are distinct from those in the minimal model.
Moreover, the tiny nonzero neutrino masses resulting from neutrino oscillation experiments imply that there might be some missing pieces in the lepton singlet sector of the SM, namely the right-handed neutrinos. 
This unsolved problem obviously shows that the SM itself is still incomplete.
From this point of view, the particle content of $SU(2)_L$ fermion doublet sector in the SM seems to be more complete than that of singlet one. 
Therefore, an extension with a vectorlike lepton doublet, corresponding to a next-to-minimal setup, has its own attraction.

%
%
In this paper, we consider an extension of the SM by adding a vectorlike lepton doublet and two complex scalars.
One of the scalar develops a non-zero vacuum expectation value, while the other scalar does not.
The former belongs to a hidden sector that does not directly couple to SM fermions.
The latter interacts with the SM and the vectorlike leptons via exotic Yukawa couplings.
This setup ensures that there is no mixing between the SM and the vectorlike leptons.
Since the new physics in this model contributes to the lepton anomalous magnetic moments, the experimental measurements of these observables impose essential constraints on the parameter space.

%
Beside the lepton anomalous magnetic moments, the new physics from the lepton sector of the model also leaves its imprints on observables measured at particle colliders.
In our study, the Large Electron-Positron Collider (LEP) and the Large Hadron Collider (LHC) data are taken into account to identify the viable parameter regions.
By using the LEP II data \cite{ALEPH:2013dgf} on the electron and muon pair production processes at the center-of-mass energy of 189 GeV, where the luminosity was highest, we will show that they play a complimentary role in addition to the general LEP lower limit for the beyond-SM charged particle mass.
%
%
%
After completing the LHC missions, a next generation of particle colliders extending the current energy and luminosity frontiers
\cite{EuropeanStrategyGroup:2020pow, 
	Adolphsen:2022ibf} 
will be necessary to explore the details of possible new physics.
To reach higher energy, one option is to use a linear collider.
Proposals of this type include
the International Linear Collider (ILC) \cite{Bambade:2019fyw}
and
the Compact Linear Collider (CLIC)
\cite{CLICdp:2018cto}.
Another option is a circular collider, of which the advantage is the small statistical uncertainties due to its large integrated luminosity.
Examples of circular $e^+e^-$ colliders are the
Circular Electron Positron Collider (CEPC)
\cite{CEPCStudyGroup:2018rmc, CEPCPhysicsStudyGroup:2022uwl},
and the
LEP3
\cite{Blondel:2012ey, Anastopoulos:2025jyh}.
While hadron colliders are the most useful for generating and detecting colored particles, lepton colliders 
\cite{FCC:2018evy, Anastopoulos:2025jyh, CEPCStudyGroup:2018rmc, 
ILC:2007bjz, 
ILC:2013jhg, 
ILC:2019gyn, 
Roloff:2018dqu, Sicking:2020gjp}
are especially sensitive to non-colored particles.
The Future Circular Collider (FCC)
\cite{FCC:2018byv, FCC:2018evy}
is one of the promising proposals since it has the advantage of both types: its first stage is an $e^+e^-$ collider (FCC-ee) and its second stage is a hadron collider (FCC-hh)
\cite{FCC:2025lpp, 
	FCC:2025uan,
	FCC:2025jtd, 
	janot_2024_yr3v6-dgh16}.
It was shown that future $e^+e^-$ colliders will be able to probe signals of new physics 
\cite{Shiltsev:2019rfl, 
	Tran:2023xzn, 
	Tran:2020tsj, Tran:2010ea, 
	Yue:2024ftz, Tran:2025azw, 
	Erdelyi:2025axy, Olgoso:2025jot
}
due to their clean environments and extremely high precision.
In our analysis, we will examine 
the ability to constrain the model's parameters at the FCC-ee.

The structure of this paper is as follows.
The model setup is described in Section 2.
In Section 3, we investigate the new physics contributions to the lepton anomalous magnetic moments and the scattering processes at
$e^+e^-$ colliders with final states of $e^+e^-$ and $\mu^+\mu^-$.
The numerical analysis is performed in detail to identify the allowed parameter space, taking into account various constraints from the LEP and the LHC data, as well as the results of lepton anomalous magnetic moment measurements.
The prospect of the FCC-ee in imposing limits on the parameter space using the muon pair production channel is discussed.
Finally, Section 4 is devoted to conclusions.






%




%

\section{Model setup}


Beside the ordinary SM particle content,
we introduce an $SU(2)_L$ doublet of vectorlike leptons with the $U(1)_Y$ charge of $\nicefrac{-1}{2}$,
\begin{align}
L_{L,R} = 
	\left(
	\begin{matrix}
		N_{L,R}	\\ E_{L,R}
	\end{matrix}
	\right),	
\end{align}
and two complex scalars $\chi$ and $\phi$ being singlets under the SM gauge symmetry.
In this model, the scalar field $\phi$ belongs to a hidden sector. It does not directly couples to SM fermions.
The interaction between the hidden and the visible sector can only be realized via the portal of scalar potential.

The Lagrangian of new physics involves 
the mass term of vectorlike leptons, the exotic Yukawa interaction term between the SM lepton, the vectorlike leptons and the new scalar field $\chi$,
and the scalar potential:
\begin{align}
\mathcal{L_\text{NP}} \supset 
	- \left( 
	m_L \overline{L_L} L_R +
	y \overline{\ell_L} L_R \chi
	+ \text{h.c.} 
	\right)
	- \mathcal{V}(\chi,\phi,H).
\label{Lagrangian}
\end{align}
The potential relevant to the new scalar fields is given in general as
\begin{align}
\mathcal{V}(\chi, \phi, H) \supset \;
&	m_\chi^2 |\chi|^2
	+ \frac{\lambda_\chi}{2} |\chi|^4
	+ m_\phi^2 |\phi|^2 
	+ \frac{\lambda_\phi}{2} |\phi|^4
	\nonumber	\\
&	+ \lambda_{\chi H} |\chi|^2 |H|^2
	+ \lambda_{\phi H} |\phi|^2 |H|^2
	+ \lambda_{\chi \phi} |\chi|^2 |\phi|^2
	+ (r \phi \chi^2 + \text{h.c.} ) ,
\label{scalarpotential}
\end{align}
where $H$ is the SM Higgs doublet.
The Lagrangian (\ref{Lagrangian}) respects a $Z_2$ symmetry, under which the SM particles and $\phi$ are even, while the vectorlike lepton doublet $L$ and $\chi$ are odd.


Among the two additional scalars, we assume that only the field $\phi$ of the hidden sector develops a non-zero vacuum expectation value (VEV), 
\begin{align}
	\left< \phi \right> =
	\sqrt{\frac{-{m'_\phi}^2}{2\lambda_\phi}} ,
\end{align}
where
\begin{align}
	{m'_\phi}^2 = m_\phi^2 + \lambda_{\phi H} 
	\left< H \right>^2.
\end{align}
Because the VEV of $\chi$ is zero, there is no mass mixing between the SM leptons and the vectorlike leptons.
After the SM Higgs field $H$ and $\phi$ acquire their VEVs, the mass term of $\chi$ can be derived from the scalar potential (\ref{scalarpotential}).
As the complex scalar field $\chi$ can be decomposed in to the real and imaginary components,
\begin{align}
	\chi = \frac{\chi_r + i\chi_i}{\sqrt{2}} ,
\end{align}
their masses are found to be
\begin{align}
	m_{\chi_r}^2 = {m'_\chi}^{2} 
	- 2r \left< \phi \right> ,	\\
	m_{\chi_i}^2 = {m'_\chi}^2 
	+ 2r \left< \phi \right> ,
\end{align}
where
\begin{align}
	{m'_\chi}^{2} = 
	m_\chi^2 + 
	\lambda_{\chi H} \left< H \right>^2 +
	\lambda_{\chi\phi} \left< \phi \right>^2	.
\end{align}
Note that the above VEV $\left< \phi \right>$ and the coupling $r$ are the sources generating a mass splitting between $\chi_r$ and $\chi_i$.
Supposing that $r>0$, $\chi_r$ is lighter than $\chi_i$. 
It becomes a stable particle due to the $Z_2$ symmetry of the theory if $m_{\chi_r} < m_L$.
As we will see latter, the mass hierarchy between $\chi_r$, $\chi_i$, and $L$ is essential when considering the LHC constraints.

\section{New physics imprints of the model}

\subsection{Lepton anomalous magnetic moments}

\begin{figure}[h!]
\begin{center}
\includegraphics[scale=1]{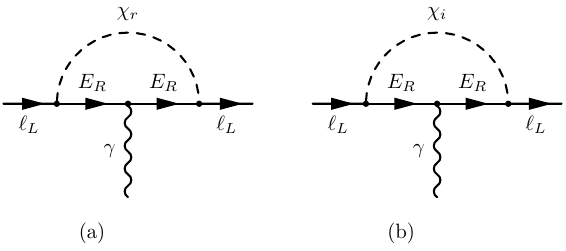}
\caption{New physics contribution to lepton anomalous magnetic moment at the leading order.}
\label{g-2}
\end{center}
\end{figure}

Due to the exotic Yuakwa interaction in the Lagrangian (\ref{Lagrangian}), the new physics contributes to the lepton anomalous magnetic moments via the loop diagrams involving the charged vectorlike lepton $E_R$ and the scalar field components $\chi_{r,i}$ as shown in Figure \ref{g-2}.
The calculated result reads \cite{Belanger:2015nma, Dinh:2020inx} 
\begin{align}
\Delta a_\ell^\text{NP} =
	\frac{y_\ell^2 m_\ell^2}{32\pi^2 m_{\chi_r}^2}
	\left[
	F_g(\tau)
	+ \frac{1}{1+\delta} 
	F_g \left(
		\frac{\tau}{1+\delta} \right)
	\right]	,
\label{a-ell}
\end{align}
where
$y_\ell$ and $m_\ell$ are the exotic Yukawa coupling and the mass of the SM lepton $\ell$, respectively.
The parameter $\tau$, $\delta$, and the loop function $F_g(x)$ are given as
\begin{align}
& \tau = \frac{m_L^2}{m_{\chi_r}^2},
	\qquad
\delta = \frac{m_{\chi_i}^2 - m_{\chi_r}^2}{m_{\chi_r}^2} ,
\\
& F_g(x) =
\frac{1}{6(1-x)^4}
\left(
6x\ln x + x^3 - 6x^2 +3x + 2
\right) .
\end{align}

\subsection{Lepton pair productions at $e^+e^-$ colliders}

\begin{figure}[h!]
\begin{center}
	\includegraphics[scale=0.9]{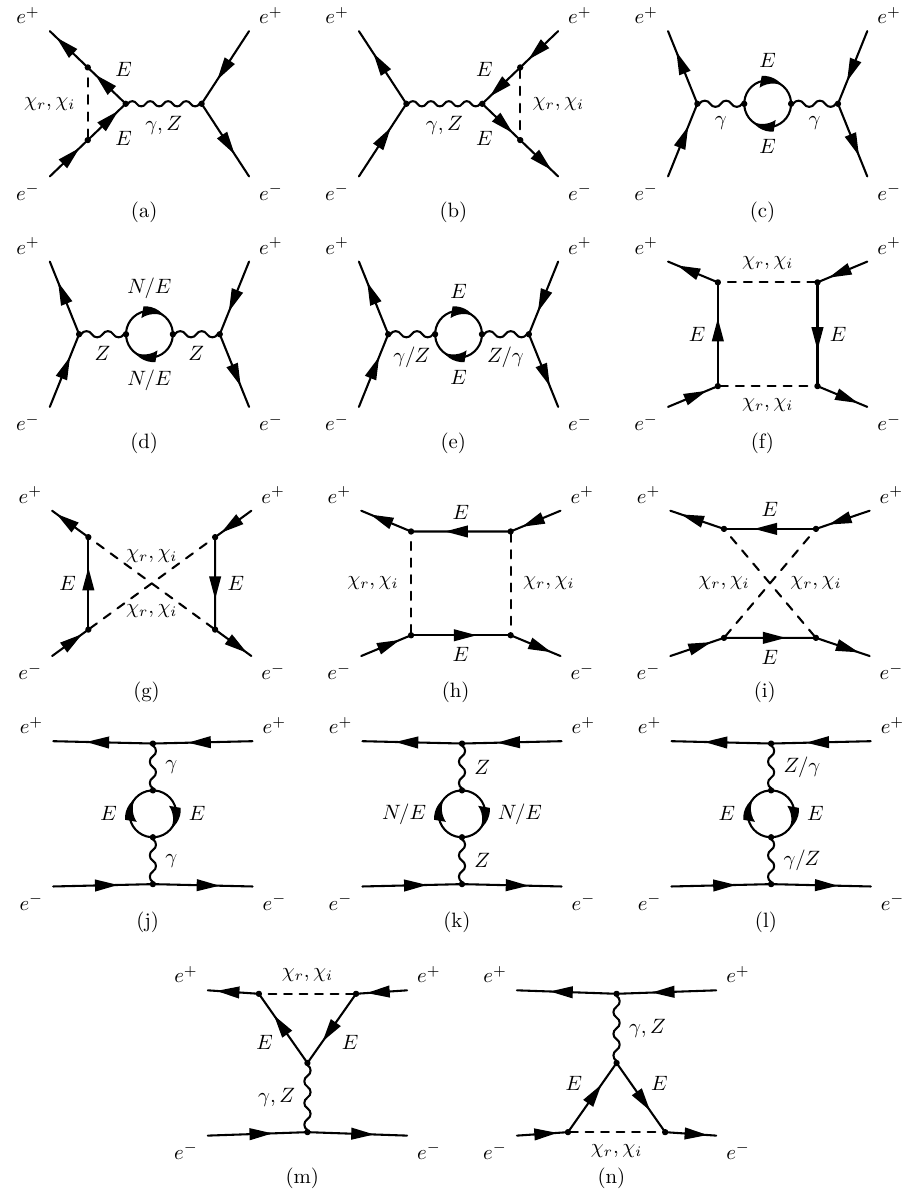}
	\caption{Feynman diagrams for the leading new physics contributions to the cross section of $e^+e^- \rightarrow e^+e^-$.}
	\label{ee2ee}
\end{center}
\end{figure}

\begin{figure}[h!]
\begin{center}
\includegraphics[scale=0.9]{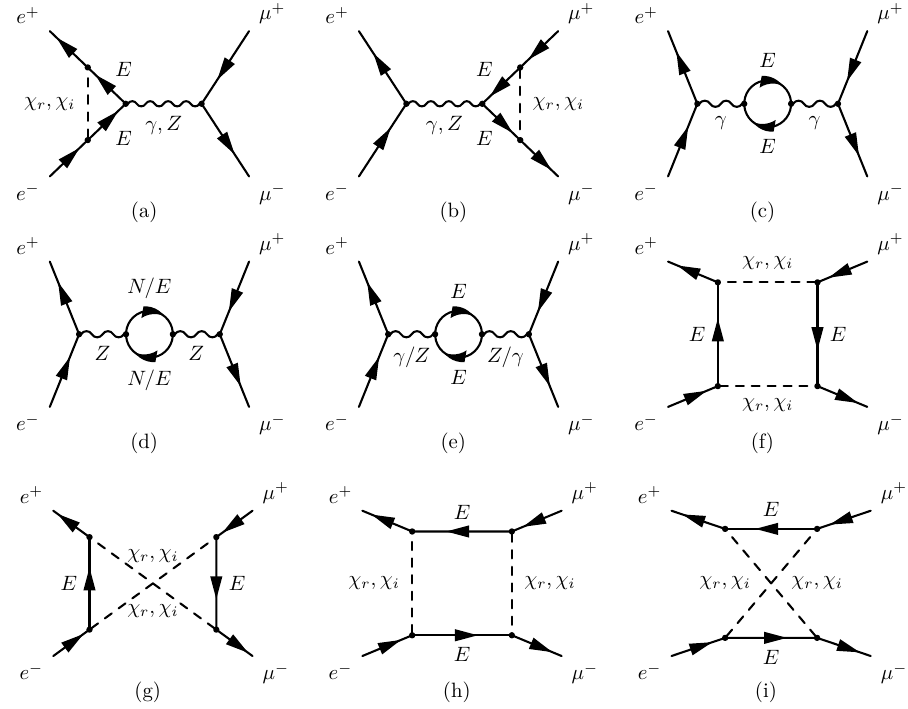}
\caption{Feynman diagrams for the leading new physics contributions to the cross section of $e^+e^- \rightarrow \mu^+\mu^-$.}
\label{ee2mm}
\end{center}
\end{figure}

Beside the lepton anomalous magnetic moments, the new particles introduced in the model also induce extra contributions to obvervables measured at particle colliders, for example scattering cross sections and forward-backward asymmetries.
Here, we consider the electron and muon pair production channels at circular $e^+e^-$ colliders.
The scattering amplitudes of these processes consist of the SM and the new physics contributions.
The Feynman diagrams for the leading new physics contribution to the scattering process 
$e^+e^- \rightarrow e^+e^-$ are depicted in Figure \ref{ee2ee}.
The diagram topologies include corrections to the $s$-channel (Figures 2a, 2b, 2c, 2d, 2e),
corrections to $t$-channels (Figures 2j, 2k, 2l, 2m, 2n), 
and box diagrams (Figures 2f, 2g, 2h, 2i).
The Feynman diagrams for the leading new physics contribution to the scattering process 
$e^+e^- \rightarrow \mu^+\mu^-$ are depicted in Figure \ref{ee2mm}.
Since this process does not have the $t$-channel at the tree level, the diagram topologies involving new physics only include 
corrections to $s$-channel (Figures 3a, 3b, 3c, 3d, 3e)
and box diagrams (Figures 3f, 3g, 3h, 3i).

In order to calculate the SM contribution to the total cross section of the process 
$e^+e^- \rightarrow e^+e^-$, 
we use the package BHWIDE version 1.06
\cite{Jadach:1995nk, Jadach:1996gu,
Placzek:1999xc, 
TwoFermionWorkingGroup:2000nks,
WorkingGrouponRadiativeCorrections:2010bjp}.
This package is a Monte Carlo event generator used for large-angle Bhabha scattering with $\mathcal{O}(\alpha)$ Yennie-Frautschi-Suura (YFS) exponentiation.
The SM cross section of the process $e^+e^- \rightarrow \mu^+\mu^-$ is calculated using the ZFITTER package version 6.42
\cite{Bardin:1999yd, Bardin:1992jc, Arbuzov:2005ma, Akhundov:2013ons}.
ZFITTER is a semi-analytical program for the calculation of observables in fermion pair production in $e^+e^-$ annihilation, taking into account complete $\mathcal{O}(\alpha)$ radiative corrections.
The new physics corrections to the two processes,
$e^+e^- \rightarrow e^+e^-$ and $\mu^+\mu^-$,
are calculated using the package
FeynCalc version 10.1.0
\cite{Mertig:1990an, Shtabovenko:2016sxi, Shtabovenko:2020gxv, Shtabovenko:2023idz}.
In this step, after writing the scattering amplitudes, we perform the tensor decomposition of one-loop integrals into Passarino-Veltman scalar integrals.
The Pasarino-Veltman functions are then evaluated numerically by
LoopTools version 2.16 \cite{Hahn:1998yk, vanOldenborgh:1989wn} 
with the help of FeynHelper version 2.0.0 \cite{Shtabovenko:2016whf}.
For the forward-backward asymmetries of these scattering processes, 
\begin{align}
	A_\text{FB} = 
	\frac{\sigma_F - \sigma_B}{\sigma_F + \sigma_B} ,
\end{align}
we calculate the cross sections in the forward and backward regions ($\sigma_{F}$ and $\sigma_B$) separately,
then substitute these quantities in the above definition.
The numerical analysis of cross sections and forward-backward asymmetries of the electron and muon pair production channels at the LEP and the FCC-ee will be presented in the next section.

\subsection{Numerical analysis}


The deviation between the measured value and the SM prediction of the muon anomalous magnetic moment is derived from Eqs. (\ref{amuEXP}) and (\ref{amuSM}):

\begin{align}
	\Delta a_\mu = a_\mu^\text{exp} - a_\mu^\text{SM}	=
	(39 \pm  64)\times 10^{-11}.
\end{align}
Similarly, the deviation between the experimental value and the SM value of the electron anomalous magnetic moment is found from Eqs. (\ref{aeEXP}) and (\ref{aeSM}) to be
\begin{align}
	\Delta a_e (\text{Rb}) = a_e^\text{exp} - a_e^\text{SM}(\text{Rb}) = 
	(34 \pm 16) \times 10^{-14} .
\end{align}
These deviations impose requirements on the sizes of new physics contributions to the anomalous magnetic moments.


Beside the lepton $g-2$, the LEP data for the eletron-positron scattering processes to 
$e^+e^-$ \cite{DELPHI:2005wxt} and $\mu^+\mu^-$ \cite{ALEPH:2013dgf}
final states
at the center of mass energy $\sqrt{s}=189$ GeV 
are used to constrain the parameter space of the model.
The luminosity at this energy (170 pb$^{-1}$) was the highest among other energy points at the LEP II experiment.
For the channel with $e^+e^-$ in the final state, according to the Ref.
\cite{DELPHI:2005wxt},
the acollinearity cut was $\theta_\text{acol} < 20^\circ$,
and
the range of scattering angle was
$\theta \in [44^\circ,136^\circ]$.
With the above acceptance cut, the cross section and the forward-backward asymmetry of this process are given as
\begin{align}
	\sigma(e^+e^-\rightarrow e^+e^-) &=
	22.73 \pm 0.46 \text{ (pb)}, \\
	A_\text{FB}(e^+e^-)	&=
	0.804 \pm 0.010 .
\end{align}
For the muon pair production process, the acceptance cut was chosen to be
$\sqrt{s'/s}>0.85$.
The cross section and the forward-backward asymmetry of this channel read \cite{ALEPH:2013dgf}
\begin{align}
	\sigma(e^+e^-\rightarrow\mu^+\mu^-) &= 
	3.150 \pm 0.077 
	\text{ (pb)},	\\
	A_\text{FB} (\mu^+\mu^-) &= 
	0.571 \pm 0.021 .
\end{align}


Regarding the future prospect at the FCC-ee, both luminosity and colliding energy will be beyond the LEP limit.
We are interested in 
the new physics effect of the model on the muon pair production cross section at the FCC-ee.
According to the FCC conceptual design report \cite{FCC:2018byv, FCC:2018evy}, there are two projected modes with $\sqrt{s}$ being 240 GeV and 365 GeV, which exceed the LEP collding energies.
In the recent FCC-ee feasibility study report \cite{FCC:2025lpp, janot_2024_yr3v6-dgh16},
the corresponding integrated luminosities at these two energy levels are proposed to be
10.8 ab$^{-1}$ and 2.7 ab$^{-1}$, respectively.
Assuming that the center values of cross section measurements are the same as the SM predictions,
and that the statistical and systematic uncertainties are equivalent \cite{Tran:2025azw},
the measured cross sections of FCC-ee are expected to be
\begin{align}
\sigma(e^+e^-\rightarrow\mu^+\mu^-) 
	= 1.89721 \pm 0.00033 \text{ (pb)},
	\quad \text{at }
	\sqrt{s}=240\text{GeV},	\\
\sigma(e^+e^-\rightarrow\mu^+\mu^-) = 0.78920 \pm 0.00042 \text{ (pb)}, 
	\quad \text{at }
	\sqrt{s}=365\text{GeV}.
\end{align}
We anticipate that such high precision measurements will have a big impact on the parameter space relevant to new physics.

In our numerical analysis, we consider the 2$\sigma$ allowed ranges for the current constraints including those from the electron and muon anomalous magnetic moments, as well as the LEP data on cross sections and forward-backward asymmetries.
Since the model contains new charged vectorlike leptons and neutral singlet scalars, its signal would be similar to the case of a supersymmetric model with charged sleptons and neutralinos in the searches for the electroweak production of these particles.
Therefore, the LHC constraints at 95\% confident level from the 
ATLAS \cite{ATLAS:2024lda}
and
CMS \cite{CMS:2024wzb}
experiments are considered in the case BR$(L\rightarrow \chi_r \ell)=100\%$.
Since the expected FCC-ee constraints are very stringent, the 3$\sigma$ allowed ranges of the muon pair production cross sections are taken into account.
In particular, we have
\begin{align}
& 1.6\times10^{-14} \leq \Delta a_e	\leq 66.0\times10^{-14} ,
	& [2\sigma]	
	\label{delta-ae}\\
& -8.9\times10^{-10} \leq \Delta a_\mu \leq 16.6\times10^{-10} ,
	& [2\sigma]
	\label{delta-amu}	\\
&21.81 \text{ pb} \leq \sigma(e^+e^-\rightarrow e^+e^-)
\leq 23.65 \text{ pb} ,
	&[189\text{ GeV},\,2\sigma]
	\label{sigma-ee}	\\
&0.783 \leq	
A_\text{FB} (e^+e^-) 
\leq	0.825 ,	
	&[189\text{ GeV},\,2\sigma]
	\label{AFB-ee}	\\
&2.996 \text{ pb}	\leq \sigma(e^+e^-\rightarrow\mu^+\mu^-) \leq	3.304 	\text{ pb},		
	&[189\text{ GeV},\,2\sigma]
	\label{sigma-mm}	\\
&0.529 \leq	
	A_\text{FB} (\mu^+\mu^-) 
	\leq	0.613 ,	
	&[189\text{ GeV},\,2\sigma]
	\label{AFB-mm}	\\
&1.89623 \text{ pb} \leq
	\sigma(e^+e^-\rightarrow\mu^+\mu^-)_\text{FCC} \leq 1.89819 \text{ pb},
	&[\text{FCC-ee, }240\text{ GeV},\,3\sigma]
	\label{FCCee-240}	\\
&0.78793 \text{ pb} \leq
	\sigma(e^+e^-\rightarrow\mu^+\mu^-)_\text{FCC} \leq 0.79047 \text{ pb} .
	&[\text{FCC-ee, }365\text{ GeV},\,3\sigma]
	\label{FCCee-365}
\end{align}
Moreover, other constraints include
\begin{align}
&	m_L \gtrsim 100 \text{ GeV},
	\label{LEPmass}	\\
&	m_L > m_{\chi_r},
	\label{stable-chir}	\\
&	y_{e,\mu} < \sqrt{4\pi},
	\label{perturbation}
\end{align}
that are due to 
the general LEP constraint on the mass of heavy charged particle \cite{ALEPH:2002nwp, 
	DELPHI:2003uqw, 
	L3:2003fyi}, 
the cosmological requirement that a new stable particle must be neutral,
and
the perturbation limit on the exotic Yukawa couplings,
respectively.
Relevant to the above constraints, the free input parameters of the model are 
the vectorlike lepton mass,
the mass of the real component of the scalar $\chi$,
the relative difference between the squared masses of the imaginary and the real components of $\chi$,
the exotic Yukawa couplings of electron and muon:
\begin{align}
\{m_L, m_{\chi_r}, \delta, 
y_e, y_\mu   \}	.
\end{align}
In our investigation, we choose $\delta=1$ as a representative benchmark.

\subsubsection{%
Benchmark investigation}

\paragraph{a. Constraints on $(m_L, m_{\chi_r})$ plane\\}

\begin{figure}[h!]
	\begin{center}
		\includegraphics[scale=0.8]{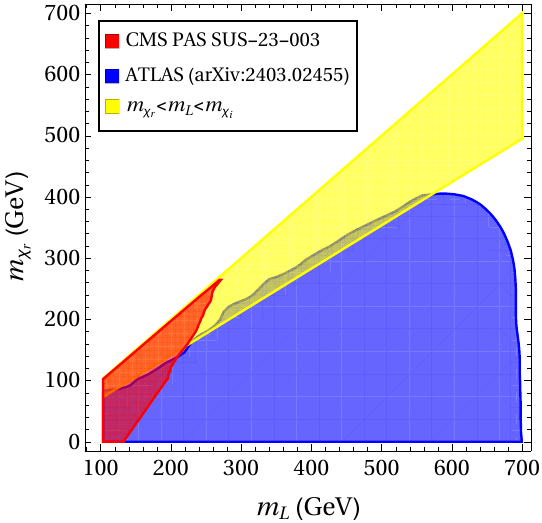}
	\caption{The LHC constraint on $(m_L,m_{\chi_r})$ plane.
		Assuming 
		BR($E^\pm \rightarrow \ell^\pm +\chi_r$)=100\%, 
		the blue and the red regions would be excluded at 95\% confident level by the ATLAS \cite{ATLAS:2024lda} and the CMS \cite{CMS:2024wzb} experiments, respectively. 
		The yellow region satisfies the condition $m_{\chi_r}<m_L<m_{\chi_i}$
		in the case $\delta=1$.}
	\label{LHC}
	\end{center}
\end{figure}

The LHC constraint on the pair of parameters $(m_L,m_{\chi_r})$ can be derived from the ATLAS and CMS search results
for the pair production of sleptons decaying into two leptons with missing energy.
In Figure \ref{LHC}, the blue region would be excluded by the ATLAS experiment \cite{ATLAS:2024lda},
while the red region would be excluded by the CMS experiment \cite{CMS:2024wzb}
at 95\% confident level.
However, in these LHC searches for supersymmetry signals, it was assumed that the slepton decays into the lightest neutralino and an SM lepton with the branching ratio of 100\%.
In our model, the charged vectorlike lepton $E$ decays to the lightest scalar $\chi_r$ and an SM charged lepton with this branching fraction only in the case 
\begin{align}
	m_{\chi_r} < m_L < m_{\chi_i}.
	\label{masshierarchy}
\end{align}
This region is highlighted with yellow color in Figure \ref{LHC} where $\delta=1$ is chosen as a representative benchmark.
In the region below the yellow area where $m_{\chi_r} < m_{\chi_i} < m_L$,
the charged vectorlike lepton $E$ decays to $\chi_r$ with the branching ratio less than 100\%. 
Hence, it is not excluded, even for the parts marked with blue or red colors below the yellow area.
For example, the point corresponding to $m_L = 140$ GeV and $m_{\chi_r}=50$ GeV in Figure \ref{LHC} is still viable.
Therefore, the actual regions excluded by the LHC results are 
the red and the blue portions that overlap with the yellow area.
In this plot, the parameter region with $m_L < 103.5$ GeV is not considered because it is excluded by the LEP run-2 experiment.
Meanwhile, the upper-left corner of the plot corresponding to $m_L < m_{\chi_r}$ is not of interest from the cosmological point of view as it predicts a stable vectorlike charged lepton.

\begin{figure}[h!]
	\begin{center}
	\includegraphics[scale=0.8]{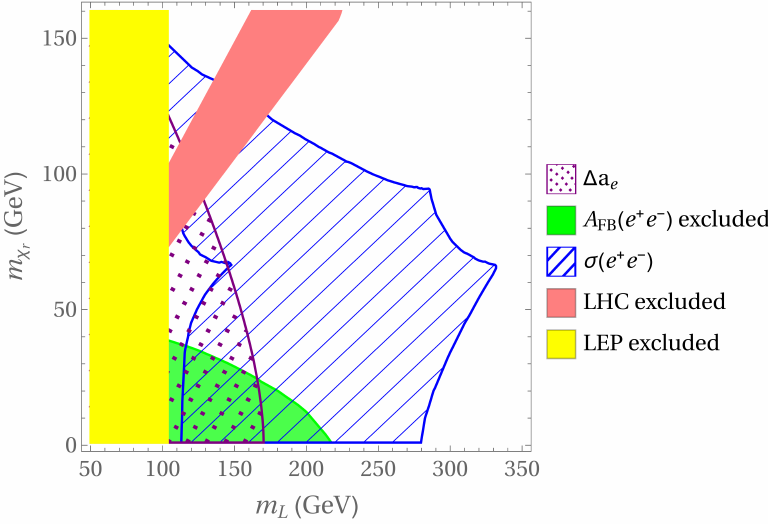}
	\caption{Constraints on $(m_L, m_{\chi_r})$ plane for $y_e=1.3$ and $y_\mu=0.12$. 
	Dotted and blue hatch regions: allowed by the constraints (\ref{delta-ae}) (\ref{sigma-ee}) on $\Delta a_e$ and $\sigma(e^+e^- \rightarrow e^+e^-)$.
	Green, pink, and yellow regions: excluded by the constraint (\ref{AFB-ee}) on $A_{FB}(e^+e^-\rightarrow e^+e^-)$, the LHC search (Figure \ref{LHC}) and the LEP data (\ref{LEPmass}).
	}
	\label{189mxml}
	\end{center}
\end{figure}
The current constraints on the $(m_L, m_{\chi_r})$ plane for $y_e=1.3$ and $y_\mu=0.12$ are shown in Figure \ref{189mxml}.
The dotted region satisfies the constraint (\ref{delta-ae}) on the electron anomalous magnetic moment.
The points with large values of $m_L$ or $m_{\chi_r}$ are not favored since they predict too small $\Delta a_e$.
Meanwhile, the whole parameter range is allowed by the constraint (\ref{delta-amu}) on $\Delta a_\mu$.
The blue hatch region is consistent with the constraint (\ref{sigma-ee}) on the Bhabha scattering cross section at $\sqrt{s}=189$ GeV.
The green region is ruled out by the constraint (\ref{AFB-ee}) on the forward-backward asymmetry of the Bhabha scattering.
The pink and yellow regions are excluded by the LHC search (Figure \ref{LHC}) and the LEP data (\ref{LEPmass}), respectively.
Our calculation shows that the constraints (\ref{sigma-mm}) and (\ref{AFB-mm}) on the cross section and the forward-backward asymmetry of the process $e^+e^- \rightarrow \mu^+\mu^-$ are satisfied for parameter points outside the yellow band in Figure \ref{189mxml}. 
In this case, the LEP data on the cross section (blue hatch region) and the forward-backward asymmetry (the green region) of the Bhabha process plays a complementary role alongside the general LEP mass constraint (\ref{LEPmass}) (the yellow region).

\begin{figure}[h!]
	\begin{center}
		\includegraphics[scale=0.8]{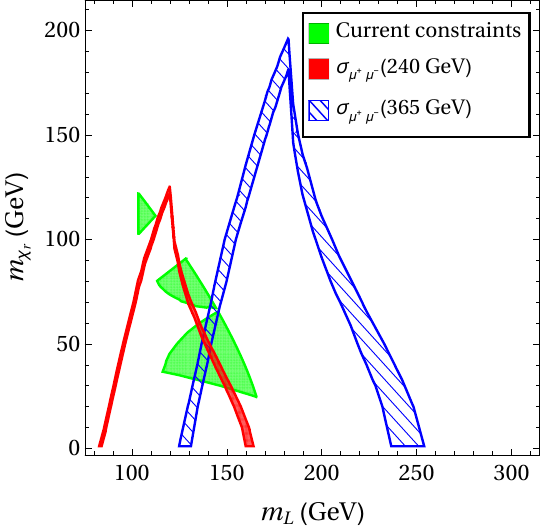}
		\caption{FCC-ee and current constraints on $(m_L, m_{\chi_r})$ plane for $y_e=1.3$ and $y_\mu=0.12$.
		The solid red and back-hatch blue regions: allowed by the projected FCC-ee data (\ref{FCCee-240})-(\ref{FCCee-365}) on $\sigma(e^+e^-\rightarrow \mu^+\mu^-)$
		at $\sqrt{s} = 240$ GeV and 365 GeV, respectively.
		The green regions are allowed by all the current constraints in Figure \ref{189mxml}.
		}
		\label{FCCmxml}
	\end{center}
\end{figure}
\begin{figure}[h!]
	\begin{center}
		\includegraphics[scale=0.8]{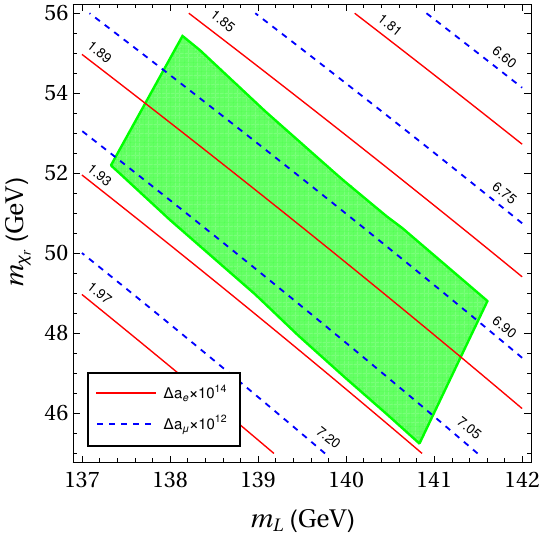}
		\caption{Combined current and expected FCC-ee constraints on $(m_L, m_{\chi_r})$ plane for $y_e=1.3$ and $y_\mu=0.12$.
		The red solid and the blue dashed lines are contours of $\Delta a_e$ and $\Delta a_\mu$.}
		\label{ALLmxml}
	\end{center}
\end{figure}

From Figure \ref{189mxml}, three disconnected regions satisfying all the considered current constraints are identified.
They are plotted separately in Figure \ref{FCCmxml} with the green color.
The red solid and the blue hatch strips are determined from the expected FCC-ee constraints (\ref{FCCee-240}) and (\ref{FCCee-365}) on the muon pair production cross section at $\sqrt{s} = 240$ GeV and 365 GeV, respectively.
These requirements are very stringent due to the high luminosity at the FCC-ee.
We observe that there exists only a tiny region satisfying both of these future constraints.
The overlapping region from Figure \ref{FCCmxml} is extracted and shown separately in Figure \ref{ALLmxml} with the green color.
Combining all the current and the expected FCC-ee constraints, the viable parameter space on ($m_L, m_{\chi_r}$) plane becomes very narrow.
The allowed range for $m_L$ is [137.3, 141.6] GeV, while that for $m_{\chi_r}$ is [45.2, 55.5] GeV.
In this figure, the contours of $\Delta a_e$ and $\Delta a_\mu$ are drawn by red solid and blue dashed lines, respectively.
With fixed values of the exotic Yukawa couplings $y_e$ and $y_\mu$, the magnitude of new physics contributions to the electron and muon anomalous magnetic moments in such small allowed mass region do not vary significantly.
We find that the average values of $\Delta a_e$ in this green region is about $1.89 \times 10^{-14}$, and
that of $\Delta a_\mu$ is about 
$6.98 \times 10^{-12}$.

\paragraph{b. Constraints on $(y_e, y_\mu)$ plane\\}


\begin{figure}[h!]
\begin{center}
	\includegraphics[scale=0.9]{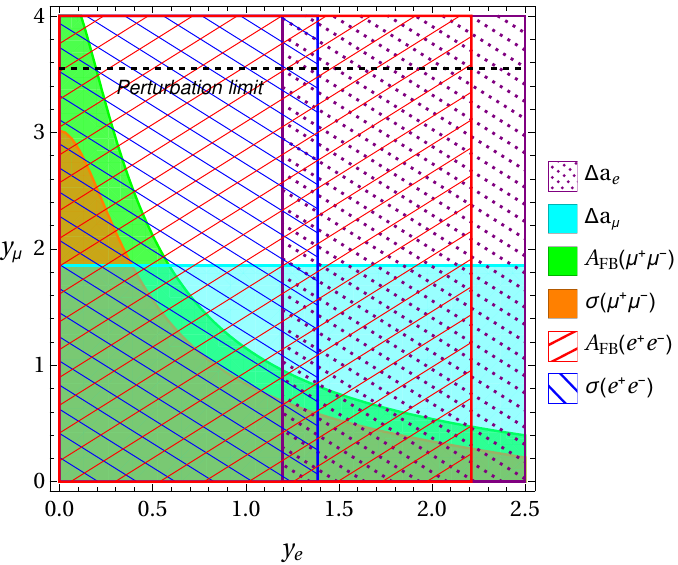}
\caption{Constraints on $(y_e,y_\mu)$ plane for $m_L=140$ GeV and $m_{\chi_r}=50$ GeV. 
Dotted and cyan regions: allowed by the constraints on $\Delta a_e$ (\ref{delta-ae}) and $\Delta a_\mu$ (\ref{delta-amu}).
Orange and green regions: allowed by the LEP data at $\sqrt{s}=189$ GeV on $\sigma(e^+e^-\rightarrow \mu^+\mu^-)$ (\ref{sigma-mm}) and $A_{FB}(e^+e^-\rightarrow \mu^+\mu^-)$ (\ref{AFB-mm}).
Blue back-hatch and red hatch regions:
allowed by the LEP data at $\sqrt{s}=189$ GeV on $\sigma(e^+e^-\rightarrow e^+e^-)$ (\ref{sigma-ee}) and $A_{FB}(e^+e^-\rightarrow e^+e^-)$ (\ref{AFB-ee}).
Dashed horizontal line indicates the perturbation limit (\ref{perturbation}) on $y_\mu$.
}
\label{189yeym}
\end{center}
\end{figure}
In Figure \ref{189yeym}, the current constraints on the parameter space of $(y_e, y_\mu)$ are plotted in the case of 
$m_L=140$ GeV and $m_{\chi_r}=50$ GeV.
The dotted and cyan regions are allowed by the constraints (\ref{delta-ae}) and (\ref{delta-amu}) on the electron and muon anomalous magnetic moments at the $2\sigma$ level.
These two conditions independently impose restrictions on the electron and muon exotic Yukawa couplings.
According to that,
the regions with $y_e$ smaller than about 1.2 or $y_\mu$ larger than about 1.8 are excluded.
The lower limit on $y_e$ 
(the upper limit on $y_\mu$) is set by the
lower bound on $\Delta a_e$ (\ref{delta-ae}) 
(the upper bound on $\Delta a_\mu$ (\ref{delta-amu})).
The electron exotic Yukawa coupling $y_e$ is further constrained by the LEP data on the cross section and the forward-backward asymmetry of the Bhabha scattering process 
$e^+e^- \rightarrow e^+e^-$,  
namely Eqs. (\ref{sigma-ee}) and (\ref{AFB-ee}).
The corresponding regions in Figure \ref{189yeym} are drawn as the blue back-hatch and the red hatch ones.
They have the form of vertical bands only restricting the range of $y_e$, because these observables are independent of $y_\mu$.
The upper bounds on $y_e$ derived from these data are approximately 1.4 and 2.2, respectively.
Meanwhile,
the LEP constraints on 
the cross section (\ref{sigma-mm})
and the forward-backward asymmetry (\ref{AFB-mm})
of the muon pair production process
are depicted in Figure \ref{189yeym} by the orange and the green regions.
The new physics contribution to the process $e^+e^- \rightarrow \mu^+\mu^-$ depends on both $y_e$ and $y_\mu$, as shown in Figure \ref{ee2mm}.
As a consequence, the allowed range for $y_\mu$ become narrower for larger values of $y_e$ and vice versa.
For both processes, the constraints from the cross sections are more severe than those from the forward-backward asymmetries.
In this case, when the general LEP requirement (\ref{LEPmass}) on $m_L$ and the LHC constraint from Figure \ref{LHC} are always satisfied, the measured cross sections, Eqs. (\ref{sigma-ee}) and (\ref{sigma-mm}), turn out to be indispensable.

\begin{figure}[h!]
\begin{center}
	\includegraphics[scale=0.8]{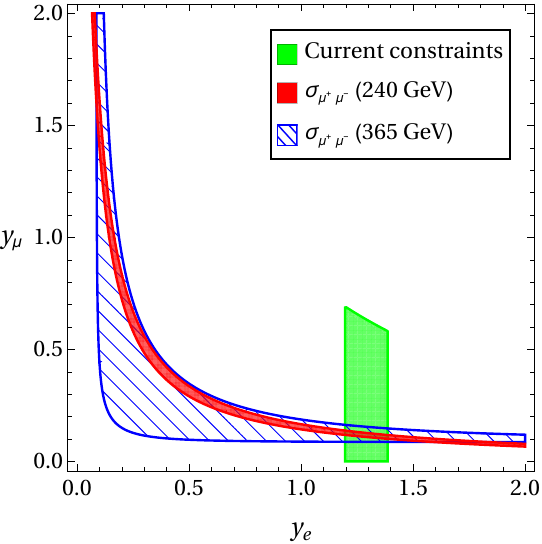}
\caption{FCC-ee and current constraints on $(y_e,y_\mu)$ plane for $m_L=140$ GeV and $m_{\chi_r}=50$ GeV.
The solid red and back-hatch blue regions: allowed by the projected FCC-ee data (\ref{FCCee-240})-(\ref{FCCee-365}) on $\sigma(e^+e^-\rightarrow \mu^+\mu^-)$
at $\sqrt{s} = 240$ GeV and 365 GeV, respectively.
The green region is allowed by all the current constraints in Figure \ref{189yeym}.
}
\label{FCCyeym}
\end{center}
\end{figure}
The expected FCC-ee constraints, Eqs. (\ref{FCCee-240}) and (\ref{FCCee-365}), 
on the muon pair production cross section at $\sqrt{s}=240$ GeV and 365 GeV lead to the solid red and the back-hatch blue regions in Figure \ref{FCCyeym}.
In this plot, the masses of the vectorlike lepton $L$ and the scalar $\chi_r$ are fixed to be the same as those in Figure \ref{189yeym}.
Since the designed FCC-ee luminosity at $\sqrt{s}=240$ GeV is larger than that at $\sqrt{s}=365$ GeV, the uncertainty of the former is smaller than the uncertainty of the latter.
It follows that the solid red region is significantly smaller than the back-hatch blue region.
We observe that the red area is mostly contained inside the back-hatch blue one, except for the case
$y_e \gtrsim 1.6$ or
$y_\mu \gtrsim 1.6$ where the former extends outside the latter.
The green band in this plot denotes the overlapping region of all the current constraints 
(\ref{delta-ae})-(\ref{AFB-mm}) extracted from Figure \ref{189yeym}.
According to that, the allowed ranges of $y_e$ and $y_\mu$ are
$[1.20, 1.39]$ and $[0, 0.7]$, respectively.
Within these allowed ranges, we see that the perturbation limits (\ref{perturbation}) on $y_e$ and $y_\mu$ are automatically satisfied.
It is clear from this figure that the expected FCC-ee will be able to exclude a significant part of the viable parameter space due to its high precision.
After imposing the FCC-ee constraints,
the expected allowed range for $y_\mu$ is approximately reduced to  $[0.101, 0.138]$.
\begin{figure}[h!]
	\begin{center}
		\includegraphics[scale=0.8]{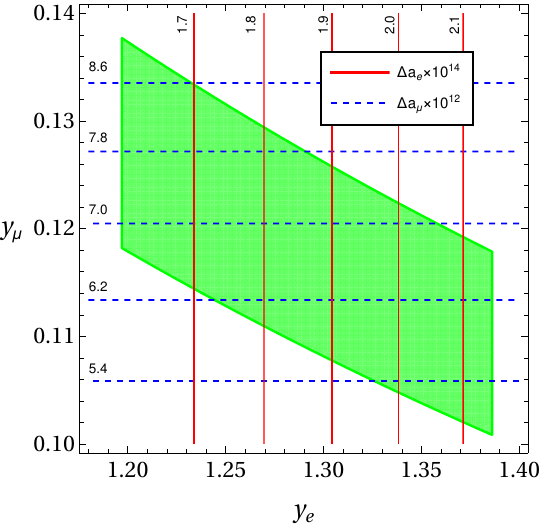}
		\caption{Combined current and expected FCC-ee constraints on $(y_e,y_\mu)$ plane for $m_L=140$ GeV and $m_{\chi_r}=50$ GeV.
			The red solid and the blue dashed lines are contours of $\Delta a_e$ and $\Delta a_\mu$.}
		\label{ALLyeym}
	\end{center}
\end{figure}
The combined region of all current and future constraints is depicted in Figure \ref{ALLyeym} as the green area.
In this figure, the contours of $\Delta a_e$ and $\Delta a_\mu$ are presented by red lines and blue dashed lines, respectively.
They are vertical and horizontal lines because
$\Delta a_e$ ($\Delta a_\mu$) only depends on $y_e$ ($y_\mu$), see Eq. (\ref{a-ell}).
In this region, the average values of the new physics contribution to the electron and muon anomalous magnetic moments are roughly
$1.9\times 10^{-14}$ and
$6.9\times 10^{-12}$.
%
%

\paragraph{c. Constraints on $(m_L,y_e)$ plane\\}

\begin{figure}[h!]
	\begin{center}
		\includegraphics[scale=0.8]{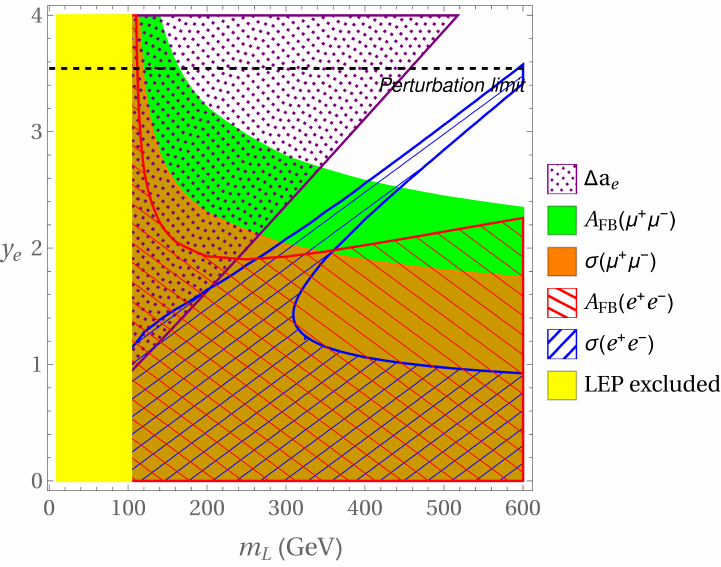}
	\caption{Constraints on $(m_L,y_e)$ plane for $m_{\chi_r}=50$ GeV and $y_\mu=0.12$. 
	The dotted region satisfies the constraint (\ref{delta-ae}) on $\Delta a_e$. 
	The orange, the green, the blue hatch, and the red back-hatch regions are allowed by the LEP constraints (\ref{sigma-mm}), (\ref{AFB-mm}), (\ref{sigma-ee}), and (\ref{AFB-ee}) at $\sqrt{s}=189$ GeV on cross sections and forward-backward asymmetries of scattering channels with the final state of $\mu^+\mu^-$ and $e^+e^-$, respectively. 
	The yellow band is excluded by LEP data (\ref{LEPmass}).
	The horizontal dashed line is the perturbation limit on $y_e$.
	}
	\label{189yeML}
	\end{center}
\end{figure}
The constraints on ($m_L, y_e$) plane are shown in Figure \ref{189yeML} with $m_{\chi_r}=50$ GeV and $y_\mu=1.2$.
The LEP exclusion on vectorlike lepton mass below about 100 GeV (\ref{LEPmass}) is represented by the yellow band.
For the whole plot ranges of this figure on the right of the yellow band, the LHC constraint from Figure \ref{LHC} and the condition (\ref{delta-amu}) on $\Delta a_\mu$ are always satisfied.
The dotted regions satisfies the constraint (\ref{delta-ae}) on $\Delta a_e$.
Combined with the perturbation limit on $y_e$ (the black dashed line),
the allowed ranges specified by these requirement for $m_L$ and $y_e$ are
approximately 
$[103.5, 460]$ GeV and 
$[0.9, \sqrt{4\pi}]$, respectively.
These ranges are further restricted once the conditions (\ref{sigma-ee})-(\ref{AFB-mm}) are taken into account.
In Figure \ref{189yeML},
the blue hatch and the red back-hatch regions satisfy the constraints on cross section and forward-backward asymmetry of the Bhabha scattering.
The orange and the green regions satisfy the constraints on cross section and forward-backward asymmetry of the muon pair production process.
Note that the orange area is plotted on top of the green one.
Applying all these constraints, the allowed parameter space on ($m_L, y_e$) plane is severely narrowed down.
The resulting viable ranges for $m_L$ and $y_e$ are 
$[103.5, 205]$ GeV and $[0.92, 1.67]$, respectively.

\begin{figure}[h!]
\begin{center}
	\includegraphics[scale=0.8]{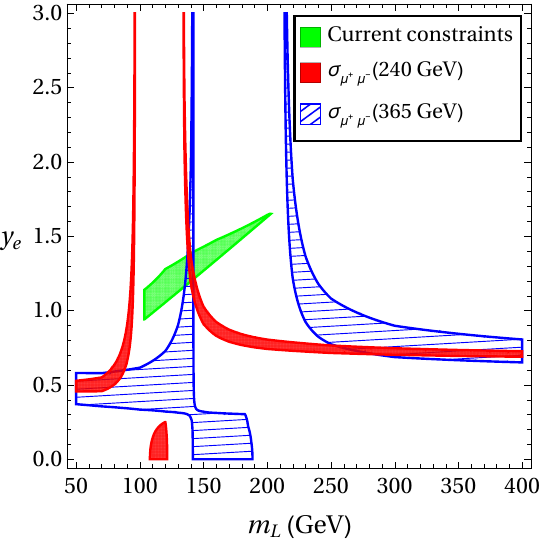}
	\caption{FCC-ee and current constraints on $(m_L,y_e)$ plane for $m_{\chi_r}=50$ GeV and $y_\mu=0.12$.
	The solid red and the blue hatch regions are determined by the expected FCC-ee results on the muon pair production cross section at $\sqrt{s}=240$ GeV and 365 GeV. The green regions are allowed by all the current constraints in Figure \ref{189yeML}.
	}
	\label{FCCyeML}
\end{center}
\end{figure}
The overlapping region determined from all the current constraints is extracted from Figure \ref{189yeML} and shown in Figure \ref{FCCyeML} as the green area.
The boundary of this area is specified by the combination of Eqs. (\ref{delta-ae}), (\ref{sigma-ee}), and (\ref{LEPmass}).
The $2\sigma$ bounds from the expected FCC-ee results of the muon pair production cross section at $\sqrt{s}=240$ GeV and 365 GeV are presented in this figure by the solid red and the blue hatch regions.
In the case $m_{\chi_r}=50$ GeV and $y_\mu=0.12$, these two expected regions overlap at three distinct ranges of the electron exotic Yukawa coupling:
(\textit{i})  $0.45 \lesssim y_e\lesssim 0.61$,
(\textit{ii}) $y_e \simeq 0.7$,
(\textit{iii}) $1.05 \lesssim y_e \lesssim 1.65$.
Putting together the current constraints (the green region), we see that only the last range of $y_e$ is acceptable.
This is mainly due to the measurement of the electron anomalous magnetic moment (\ref{delta-ae}) that requires relatively large values of $y_e$.
Thanks to the high luminosity of the FCC-ee, precise measurements of the cross section will be able to single out narrow viable ranges of $m_L$ and $y_e$ to be about $[137.9, 142]$ GeV and $[1.195,1.385]$.
\begin{figure}[h!]
	\begin{center}
		\includegraphics[scale=0.8]{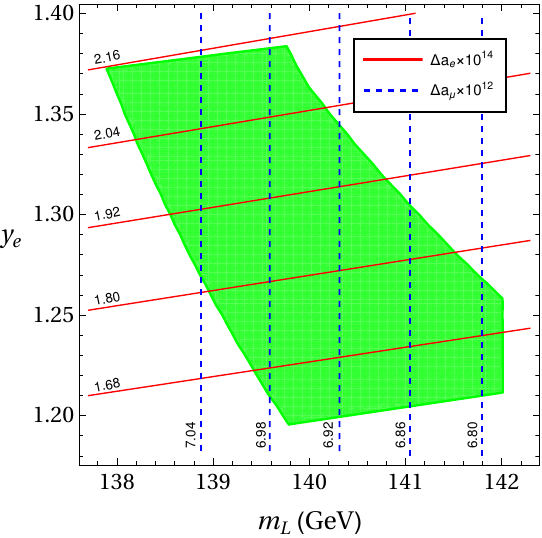}
	\caption{Combined current and expected FCC-ee constraints on $(m_L,y_e)$ plane for $m_{\chi_r}=50$ GeV and $y_\mu=0.12$.
	The red solid and the blue dashed lines are contours of $\Delta a_e$ and $\Delta a_\mu$.}
	\label{ALLyeML}
\end{center}
\end{figure}
Combining the current and the FCC-ee constraints, the overlapping region from Figure \ref{189yeML} is obtained and plotted separately in Figure \ref{ALLyeML} as the green area.
Here, the contour plots of the new physics contributions to the electron and muon anomalous magnetic moments (the red and blue dashed lines) are drawn.
While $\Delta a_e$ depends on both $m_L$ and $y_e$, $\Delta a_\mu$is independent of $y_e$ (see Eq. (\ref{a-ell})).
Therefore, the contours of $\Delta a_\mu$ are vertical lines.

\paragraph{d. Constraints on $(m_L,y_\mu)$ plane\\}

\begin{figure}[h!]
	\begin{center}
		\includegraphics[scale=0.8]{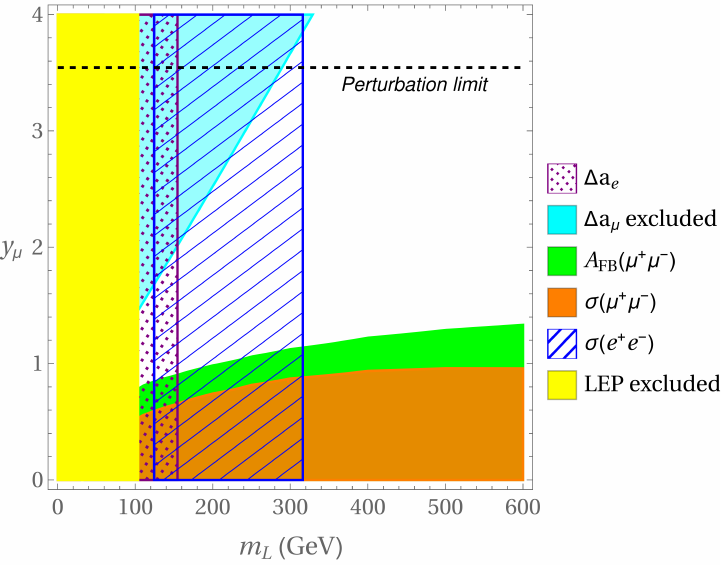}
	\caption{Constraints on $(m_L,y_\mu)$ plane for $m_{\chi_r}=50$ GeV and $y_e=1.3$.
		The dotted region is allowed by the constraint (\ref{delta-ae}) on $\Delta a_e$. The cyan region is excluded by the constraint (\ref{delta-amu}) on $\Delta a_\mu$.
	The orange and green regions satisfy the LEP data (\ref{sigma-mm})-(\ref{AFB-mm}) on cross section and forward-backward asymmetry of the muon pair production process.
	The blue hatch region satisfies the condition (\ref{sigma-ee}) on the Bhabha scattering cross section.
	The yellow band is excluded by the LEP constraint \ref{LEPmass} on $m_L$. The horizontal black dashed line represents the perturbation limit on $y_\mu$.
	}
	\label{189ymML}
\end{center}
\end{figure}
In Figure \ref{189ymML}, the current constraints are plotted on ($m_L, y_\mu$) in the case
$m_{\chi_r}=50$ GeV and $y_e=1.3$.
The LEP excluded region on the vectorlike lepton mass is shown in yellow, similar to that in Figure \ref{189yeML}.
For all the paramter points on the right-hand side of the yellow region, the LHC constraint derived from Figure \ref{LHC} is always satisfied.
The dotted region is allowed by the constraint (\ref{delta-ae}) on the electron anomalous magnetic moment.
It is a vertical band because $\Delta a_e$ does not depend on $y_\mu$.
In order to predict sizable contribution to the electron magnetic moment, the vectorlike lepton should not be too heavy.
For this benchmark, the viable range of the vectorlike lepton mass dictated by Eqs. (\ref{delta-ae}) and (\ref{LEPmass}) is 
$100 \text{ GeV} \lesssim m_L \lesssim 160$ GeV.
In this figure, the cyan region is excluded by the constraint (\ref{delta-amu}) on the muon anomalous magnetic moment.
Its boundary depends on both parameters.
The upper left corner of the plot is ruled out, since larger values of $y_\mu$ and smaller values of $m_L$ lead to too large new physics contribution (\ref{a-ell}) to the muon $g-2$.
The areas consistent with the constraints (\ref{sigma-mm}) and (\ref{AFB-mm}) on the cross section and the forward-backward asymmetry of the scattering process 
$e^+e^- \rightarrow \mu^+\mu^-$ at $\sqrt{s}=189$ GeV are depicted in Figure \ref{189ymML} with the orange and green colors, respectively.
It is observed that large $y_\mu$ is not favorable, because it predict the cross section larger than the corresponding upper bound.
Hence, the points inside the orange and the green regions satisfy the perturbation limit on $y_\mu$ (the horizontal black dash line).
In this case, the constraint on the cross section is more severe than that on the forward-backward asymmetry. 
Therefore, the orange region is contained within the green one.
Furthermore, the cyan region is well above the orange one, implying that the constraint on the muon pair production cross section is also more stringent than the one on the muon $g-2$.
For the Bhabha scattering process, the forward-backward asymmetry always satisfies the constraint (\ref{AFB-ee}) for the whole plot ranges fulfilling the LEP condition (\ref{LEPmass}) on $m_L$. 
The region allowed by the constraint (\ref{sigma-ee}) on the Bhabha cross section is marked as the blue hatch area.
It has the form of a vertical band because the Feynman diagrams contributing to this process (see Figure \ref{ee2ee}) is irrelevant to the muon exotic Yukawa coupling $y_\mu$.
As required by Eq. (\ref{sigma-ee}), the ranges
$m_L \lesssim 123.7$ GeV and 
$m_L \gtrsim 320$ GeV are ruled out. 
For this benchmark, this exclusion is more severe than the requirement in Eq. (\ref{LEPmass}).
Taking into account all the above constraints,
the viable parameter region in the plance $(m_L, y_\mu)$ is determined mainly from 
the requirements on the electron and muon pair production cross section and the measurement of electron $g-2$.

\begin{figure}[h!]
	\begin{center}
		\includegraphics[scale=0.8]{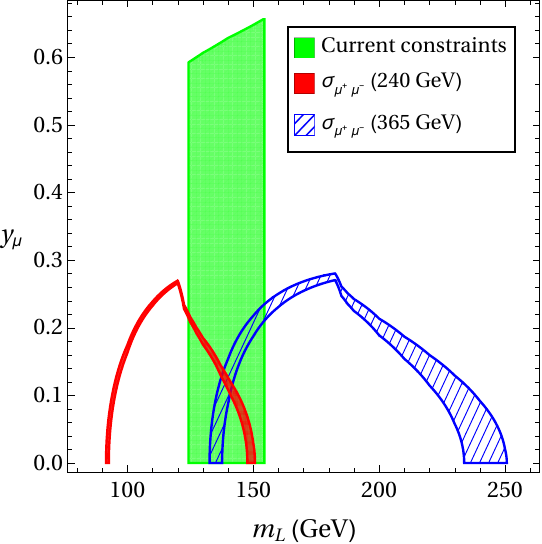}
	\caption{FCC-ee and current constraints on $(m_L,y_\mu)$ plane  for $m_{\chi_r}=50$ GeV and $y_e=1.3$. The green band satisfies all the current considered constraints (\ref{delta-ae})-(\ref{AFB-mm}). The red and blue hatch regions are consistent with the expected FCC-ee constraints (\ref{FCCee-240}) and (\ref{FCCee-365}), respectively.
	}
	\label{FCCymML}
\end{center}
\end{figure}
Superimposing all the individual areas in Figure \ref{189ymML}, the combined allowed region can be extracted and is shown in Figure \ref{FCCymML} with the green color.
The currently allowed ranges for $m_L$ and $y_\mu$ are 
[123.7, 155] GeV and [0, 0.66], respectively.
The expected FCC-ee constraints
(\ref{FCCee-240}) and (\ref{FCCee-365}) at the colliding energies of 240 GeV and 365 GeV are depicted by the red and the blue hatch strips.
The maximum values of $y_\mu$ range derived from these constraints are about 0.273 attained at 
$m_L = 120$ GeV for $\sqrt{s}=240$ GeV, and 0.282 at
$m_L = 182.5$ GeV for $\sqrt{s}=365$ GeV.
These peaks correspond to the cases where the colliding energy is just enough to generate a pair of vectorlike leptons, namely $m_L = \frac{\sqrt{s}}{2}$.
For these points, the largest new physics contribution to the cross section comes from the box diagrams in Figures \ref{ee2mm}h and \ref{ee2mm}i.
Above these peaks, the cross section is always outside its expected FCC-ee $3\sigma$ range.
Due to the high precision at the FCC-ee, the red and blue hatch strips are very thin.
Therefore, they specify an extremely limited viable region, excluding most part of the green area.

\begin{figure}[h!]
	\begin{center}
		\includegraphics[scale=0.8]{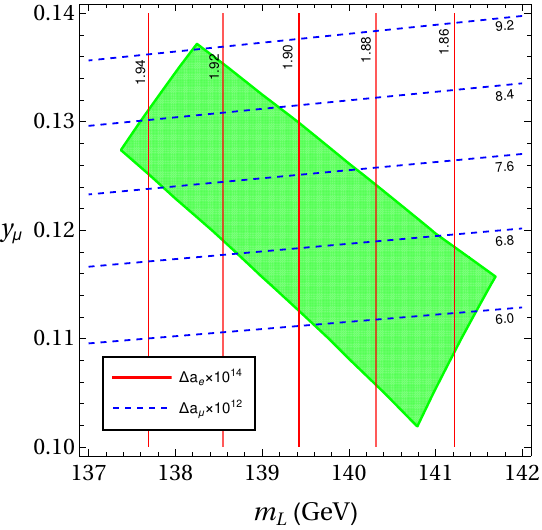}
	\caption{Combined current and expected FCC-ee constraint on $(m_L,y_\mu)$ plane for $m_{\chi_r}=50$ GeV and $y_e=1.3$. The red solid and the blue dashed lines are contours of $\Delta a_e$ and $\Delta a_\mu$.}
	\label{ALLymML}
	\end{center}
\end{figure}
The combined current and expected FCC-ee constraint is shown in Figure \ref{ALLymML} with green color.
The allowed ranges for $m_L$ and $y_\mu$ are significantly reduced to about
[137.4, 141.7] GeV and 
[0.102, 0.137], respectively.
In this figure, the contours of the new physics contributions to the electron and muon anomalous magnetic moments are presented as the red solid and the blue dashed lines.
The former are vertical because $\Delta a_e$ is independent of $y_\mu$.
Around the center of this region, the average predicted value of $\Delta a_e$ is about
$1.9 \times 10^{-14}$,
while that of $\Delta a_\mu$ is about
$7.0 \times 10^{-12}$.

\paragraph{e. Constraints on 
$(m_{\chi_r}, y_e)$ plane\\}

\begin{figure}[h!]
\begin{center}
	\includegraphics[scale=0.8]{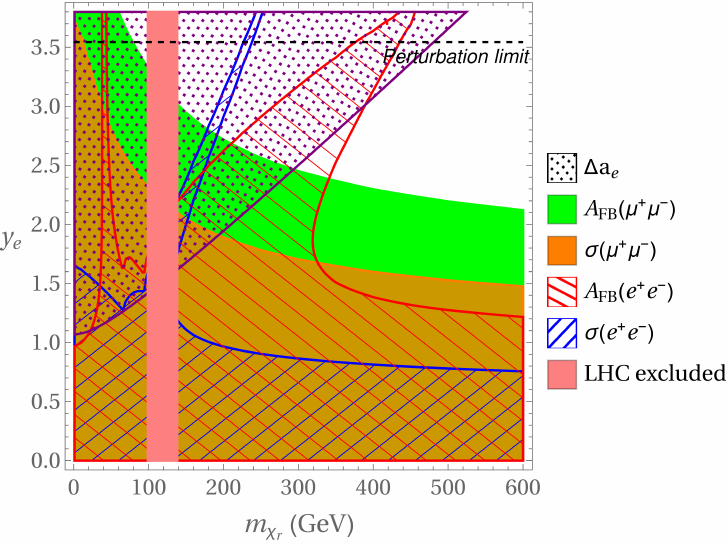}
	\caption{Constraints on $(m_{\chi_r}, y_e)$ plane for $m_L=140$ GeV and $y_\mu=0.12$. The dotted, the green, the orange, the red back-hatch, the blue hatch regions satisfies the constraints on the electron anomalous magnetic moment (\ref{delta-ae}), the forward-backward asymmetry (\ref{AFB-mm}) and the cross section (\ref{sigma-mm}) of muon pair production process, the forward-backward asymmetry (\ref{AFB-ee}) and the cross section (\ref{sigma-ee}) of Bhabha scattering process, respectively.
	The pink vertical bar is excluded by the LHC data (Figure \ref{LHC}). The horizontal black dashed line represents the perturbation limit on $y_e$.}
	\label{189yeMx}
\end{center}
\end{figure}
In Figure \ref{189yeMx}, we plot all the regions allowed by the current constraints on the plane $(m_{\chi_r}, y_\mu)$ for 
$m_L = 140$ GeV and $y_\mu=0.12$.
For this benchmark, the general LEP mass condition (\ref{LEPmass}) is always satisfied.
We find that the constraint (\ref{delta-amu}) on muon anomalous magnetic moment is satisfied over the entire plot range in this case.
The constraint (\ref{delta-ae}) requires relatively large new physics contribution to the electron magnetic moment.
Hence, it excludes the region with small $y_e$ or large $m_{\chi_r}$ (i.e. the lower right part of the figure).
The region satisfies this requirement is marked by the dotted area.
The regions consistent with the constraints on the cross section and the forward-backward asymmetry of the process $e^+e^-\rightarrow \mu^+\mu^-$ at $\sqrt{s}=189$ GeV are shown as orange and green areas, where the former is plotted on top of the latter.
We see that the constraint (\ref{sigma-mm}) is more severe than (\ref{AFB-mm}).
For the Bhabha process 
$e^+e^- \rightarrow e^+e^-$,
the regions satisfy the constraints (\ref{sigma-ee}) and (\ref{AFB-ee}) on the cross section and the forward-backward asymmetry are denoted by the blue hatch and the red back-hatch areas.
For $m_L = 140$ GeV, the LHC constraint from Figure \ref{LHC} rules out the parameter range 
100 GeV $\lesssim m_{\chi_r} \lesssim$ 140 GeV.
The corresponding excluded region is depicted by a pink vertical bar in Figure \ref{189yeMx}. 
The horizontal black dashed line represents the perturbation limit (\ref{perturbation}) on $y_e$.
For this benchmark, the combination of other constraints is enough to restrict $y_e$ to be smaller this limit.

\begin{figure}[h!]
	\begin{center}
		\includegraphics[scale=0.8]{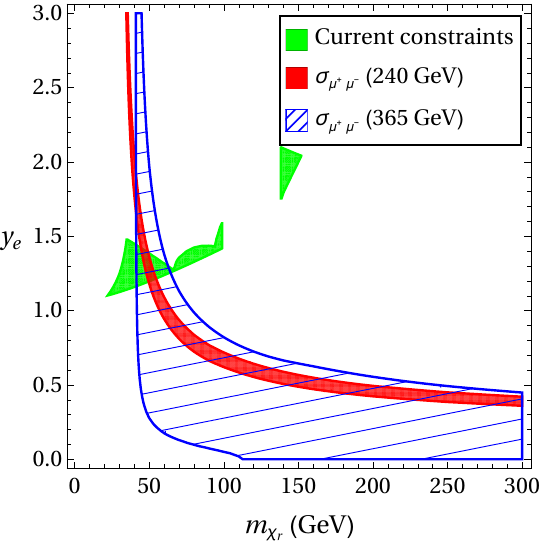}
	\caption{FCC-ee and current constraints on $(m_{\chi_r}, y_e)$ plane for $m_L=140$ GeV and $y_\mu=0.12$. The color codes are the same as those in Figure \ref{FCCymML}.}
	\label{FCCyeMx}
	\end{center}
\end{figure}
Superimposing all the allowed regions in Figure \ref{189yeMx}, we obtained two discontinuous green areas in Figure \ref{FCCyeMx} that are consistent with the set of current constraints (\ref{delta-ae})-(\ref{AFB-mm}).
Taking into account the constraint (\ref{stable-chir}), the green area on the right with $m_{\chi_r} > 140$ GeV is excluded, while the one on the left is preferable.
The expected FCC-ee constraints (\ref{FCCee-240}) and (\ref{FCCee-365}) at the colliding energies of 240 GeV and 365 GeV are represented by the red and the blue hatch strips, respectively.
The magnitude of new physics contribution to $\sigma(e^+e^- \rightarrow \mu^+\mu^-)$ becomes smaller for heavier $\chi_r$ or smaller $y_e$.
Therefore, for smaller values of $m_{\chi_r}$, larger $y_e$ is necessary to keep the cross section staying in the allowed range.
This explains the shape of the red and the blue hatched strips.
Applying the constraints (\ref{FCCee-240}) and (\ref{FCCee-365}), the separated region with $\chi_r >140$ GeV is ruled out.
Thus, we arrive at the same conclusion as using the constraint (\ref{stable-chir}).
Moreover, the future constraints significantly narrow down the green region on the left.
In particular, the viable range of $m_{\chi_r}$ is reduced from 
[20.5, 100] GeV to [44.4, 54.6] GeV,
and the viable range of $y_e$ is reduced from
[1.095, 1.605] to [1.195, 1.423].
\begin{figure}[h!]
	\begin{center}
		\includegraphics[scale=0.8]{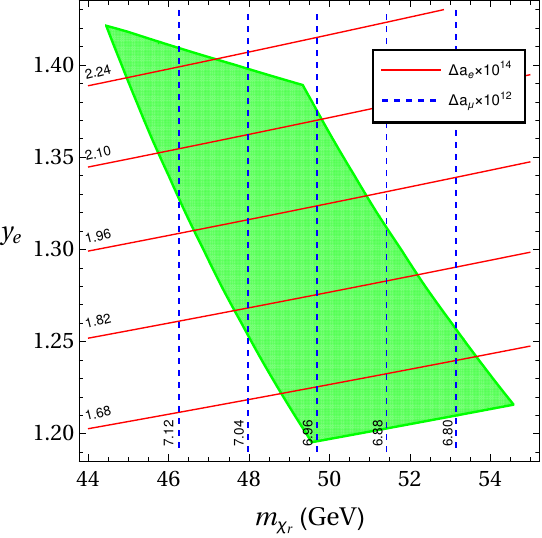}
	\caption{Combined current and expected FCC-ee constraints on $(m_{\chi_r}, y_e)$ plane for $m_L=140$ GeV and $y_\mu=0.12$. The red solid and the blue dashed lines are contours of $\Delta a_e$ and $\Delta a_\mu$.}
	\label{ALLyeMx}
\end{center}
\end{figure}
In Figure \ref{ALLyeMx}, we plot the overlapping region from Figure \ref{FCCyeMx} that satisfies all the current constraints as well as the expected FCC-ee constraints as the green area.
The red solid and the blue dashed lines are contours of $\Delta a_e$ and $\Delta a_\mu$.
Since $\Delta a_\mu$ is independent of $y_e$, the blue dashed contours are vertical.
Because the allowed mass range for $m_{\chi_r}$ is relatively small, the values of $\Delta a_\mu$ do not change drastically.
Meanwhile, $\Delta a_e$ depends on both $m_{\chi_r}$ and $y_e$.
Its value varies more along $y_e$ direction than $m_{\chi_r}$ direction.
We find that the average values of $\Delta a_e$ and $\Delta a_\mu$ around the center of this allowed region are about 
$1.89 \times 10^{-14}$ and 
$6.96 \times 10^{-12}$.

\paragraph{f. Constraints on $(m_{\chi_r}, y_\mu)$ plane\\}

\begin{figure}[h!]
\begin{center}
	\includegraphics[scale=0.8]{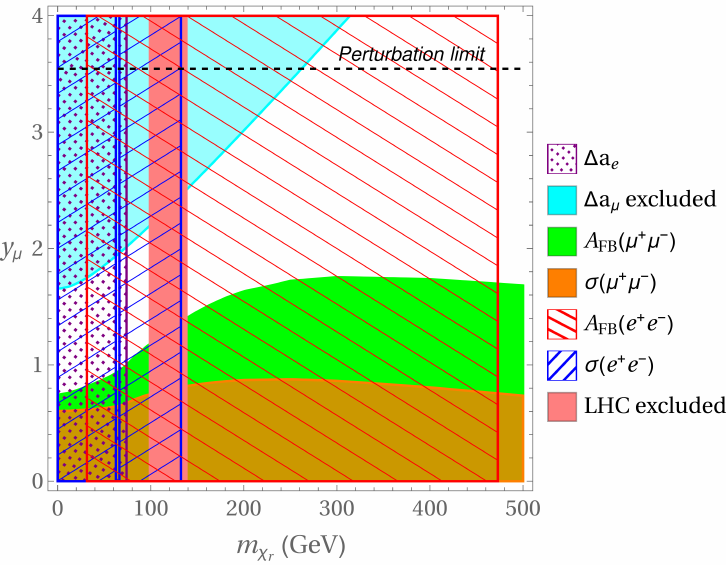}
	\caption{Constraints on $(m_{\chi_r}, y_\mu)$ plane for $m_L=140$ GeV and $y_e=1.3$. The dotted region satisfies the constraint (\ref{delta-ae}). The cyan region is ruled out by Eq. (\ref{delta-amu}).
	The orange, the green, the blue hatch, and the red back-hatch regions are allowed by the constraints (\ref{sigma-mm}), (\ref{AFB-mm}), (\ref{sigma-ee}), and (\ref{AFB-ee}), respectively. The pink vertical bar is excluded by the LHC data (Figure \ref{LHC}). The horizontal black dashed line represents the perturbation limit on $y_\mu$.}
	\label{189ymuMx}
\end{center}
\end{figure}
In Figure \ref{189ymuMx}, the current constraints are shown on $(m_{\chi_r}, y_\mu)$ plane for the case $m_L=140$ GeV and $y_e=1.3$.
The regions satisfy the constraints (\ref{delta-ae}), (\ref{sigma-ee}), and (\ref{AFB-ee}) on 
$\Delta a_e$, 
$\sigma (e^+e^- \rightarrow e^+e^-)$, and 
$A_\text{FB} (e^+e^- \rightarrow e^+e^-)$
are depicted as 
the dotted, the blue hatch, and the red back-hatch bands, correspondingly.
These bands are vertical because the relevant observables are independent of the muon exotic Yukawa coupling $y_\mu$.
Especially, the measurement on the Bhabha cross section (the blue hatch region) indicates that $m_{\chi_r}$ could not be larger than about 130 GeV.
Therefore, the condition (\ref{stable-chir}) is automatically fulfilled in this case.
The cyan region is excluded due to the constraint (\ref{delta-amu}) on the muon anomalous magnetic moment that prefers larger $m_{\chi_r}$ and smaller $y_\mu$. 
This ensures the theoretical prediction of the muon $g-2$ is within the experimental range, which is in agreement with the SM value.
The orange and the green regions are consistent with the LEP measurement results, Eqs. (\ref{sigma-mm}) and (\ref{AFB-mm}), on the cross section and the forward-backward asymmetry of the muon pair production process at $\sqrt{s}=189$ GeV.
These regions occupy the lower part of the figure, showing that only small values of $y_\mu$ is acceptable.
Hence, the perturbation limit (\ref{perturbation}) on $y_\mu$ is also fulfilled inside these regions.
Similar to Figure \ref{189yeMx}, the pink area in this figure is ruled out by the LHC constraint derived from Figure \ref{LHC}.
There are two discontinuous regions satisfying all the current constraints in Figure \ref{189ymuMx}.
They are determined solely from the constraints on the cross section and the forward-backward asymmetry of the Bhabha scattering, the muon pair production cross section, and the electron anomalous magnetic moment.

\begin{figure}[h!]
\begin{center}
	\includegraphics[scale=0.8]{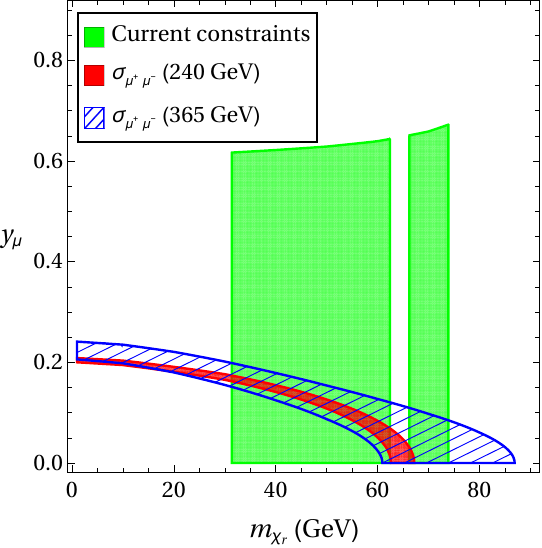}
	\caption{FCC-ee and current constraints on $(m_{\chi_r}, y_\mu)$ plane for $m_L=140$ GeV and $y_e=1.3$. The color codes are the same as those in Figure \ref{FCCymML}.}
	\label{FCCymuMx}
\end{center}
\end{figure}
These two allowed regions are plotted in Figure \ref{FCCymuMx} with the green color.
The expected FCC-ee constraints on 
$\sigma(e^+e^- \rightarrow \mu^+\mu^-)$ at the center-of-mass energies of 240 GeV and 365 GeV (Eqs. (\ref{FCCee-240}) and (\ref{FCCee-365})) are shown in this figure by the red and the blue hatch areas.
Due to the interplay between two input parameters $m_{\chi_r}$ and $y_\mu$, once the former is reduced, the latter must be increased accordingly to keep the cross sections staying within the allowed ranges.
In this case, the constraint at $\sqrt{s}=240$ GeV is more severe than the one at $\sqrt{s}=365$ GeV.
Hence, the red region is thinner and almost inside the blue hatch one.
We observe that, thanks to the high luminosity at the FCC-ee, it is able to narrow down the viable parameter space significantly.
In particular, the allowed range of $y_\mu$ is reduced from 
[0, 0.675] to 
[0, 0.174].
For the parameter $m_{\chi_r}$, there are two distinct allowed ranges.
While the left one is still the same,
the right one is reduced drastically from 
[66.1, 74.1] GeV to [66.1, 67.3] GeV due to the expected FCC-ee results.

\begin{figure}[h!]
\begin{center}
	\includegraphics[scale=0.8]{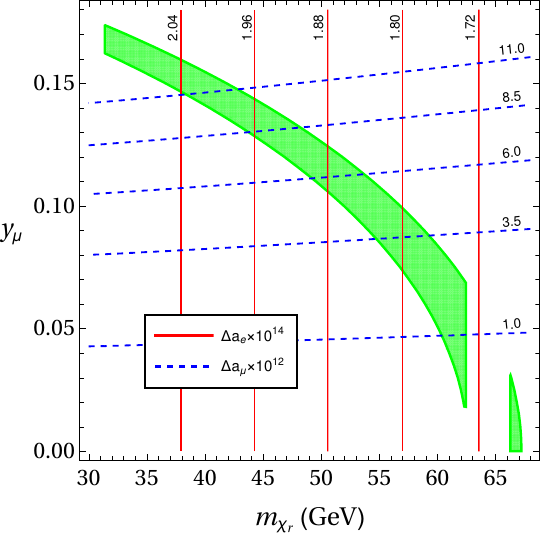}
	\caption{Combined current and expected FCC-ee constraints on $(m_{\chi_r}, y_\mu)$ plane for $m_L=140$ GeV and $y_e=1.3$. The red solid and the blue dashed lines are contours of $\Delta a_e$ and $\Delta a_\mu$.}
	\label{ALLymuMx}
\end{center}
\end{figure}

The overlapping regions satisfying all the current and the expected FCC-ee constraints are plotted in Figure \ref{ALLymuMx}.
Here, the contours of $\Delta a_e$ and $\Delta a_\mu$ are shown by red solid and blue dashed lines respectively.
The red contours are vertical because $\Delta a_e$ is independent of $y_\mu$ (see Eq. (\ref{a-ell})).
In these allowed regions, the new physics contribution to the electron $g-2$ does not change much since the viable ranges of $m_{\chi_r}$ are relatively small.
Meanwhile,  the new physics contribution to the muon $g-2$ varies by an order of magnitude in the direction close to the $y_\mu$ axis due to its strong dependence on $y_\mu$.

\subsubsection{%
Global  investigation}

A global analysis is performed by scanning over all four input parameters of the model.
We use a fixed grid for the mass pair ($m_L, m_{\chi_r}$), while letting $y_{e,\mu}$ pick up random values.
The ranges of input parameters are chosen to be as follows
\begin{align}
&	103.5 \text{ GeV} < m_L < 350 \text{ GeV} ,	\\
&	0 \text{ GeV} < m_{\chi_r} < 250 \text{GeV},	\\
&	0 < y_e < \sqrt{4\pi} ,	\\
&	0 < y_\mu < \sqrt{4\pi}	.
\end{align}
\begin{figure}[h!]
	\begin{center}
		\includegraphics[scale=0.7]{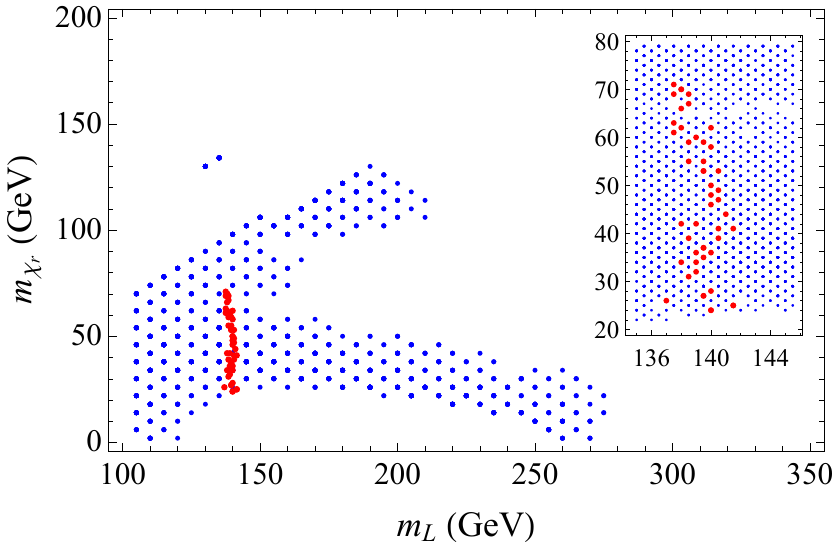}
		\caption{
		Projection of parameter points on $(m_L, m_{\chi_r})$ plane. 
		The blue points are allowed by current constraints. The red points satisfy the combined current and expected FCC-ee constraints.
		}
		\label{FullScan_mLmX}
	\end{center}
\end{figure}

In Figure \ref{FullScan_mLmX}, the allowed parameter points resulting from the full scanning are projected on ($m_L,m_{\chi_r}$) plane, and represented by blue dots.
The allowed ranges of $m_L$ and $m_{\chi_r}$, when letting $y_{e,\mu}$ free, are about 
[103.5, 275] GeV 
and [0, 135] GeV.
The lower and upper bounds of $m_L$ are determined by the LEP constraint (\ref{LEPmass}) and the data on $\Delta a_e$ (\ref{delta-ae}), respectively.
This fact can be seen in Figures (\ref{189mxml}), (\ref{189yeML}), and (\ref{189ymML}).
For $m_{\chi_r}$, its upper bound is determined by the constraint on $\Delta a_e$ (\ref{delta-ae}), as shown in Figures \ref{189mxml}, \ref{189yeMx}, and \ref{189ymuMx}.
Values of $m_L$ and $m_{\chi_r}$ larger than their upper bounds are not favored since they predict too small electron anomalous magnetic moment.
Due to the effect of varying exotic Yukawa couplings $y_{e,\mu}$, the region marked by blue dots in Figure \ref{FullScan_mLmX} is an extension of the green regions in Figure \ref{FCCmxml}.
The high precision of cross section measurement at the FCC-ee will strongly restrict the input parameter space.
For that reason, a finer scan is performed covering the entire region satisfying the combination of current data and the expected FCC-ee constraints (\ref{FCCee-240}) and (\ref{FCCee-365}) on the cross section of the muon pair production process.
The result is presented in the inset plot in Figure \ref{FullScan_mLmX}, where the red dots satisfy both current and future constraints.
These red dots are then shown in the main plot of the figure.
They form a thin vertical strip 
with $m_{\chi_r}$ spanning from 24 GeV to 71 GeV, and $m_L$ from 137 GeV to 143 GeV.
Thus, due the sensitivity of muon pair production cross section with respect to $m_L$ involving in the virtual correction and the expected highly accurate measurement at the FCC-ee, this future collider will be able to specify a very narrow range of vectorlike lepton mass $m_L$.

\begin{figure}[h!]
	\begin{center}
		\includegraphics[scale=0.7]{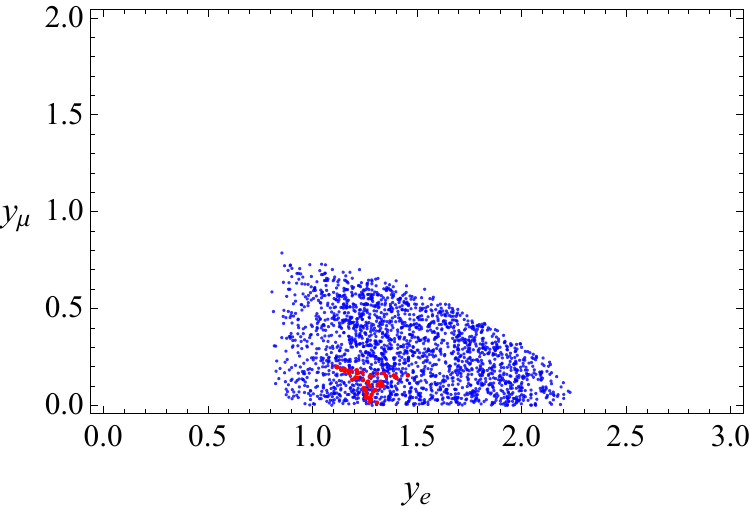}
		\caption{
		Projection of parameter points on $(y_e, y_\mu)$ plane. 
		The blue points are allowed by current constraints. The red points satisfy the combined current and expected FCC-ee constraints.
		}
		\label{FullScan_yeym}
	\end{center}
\end{figure}

The projection of parameter points on ($y_e, y_\mu$) plane is depicted in Figure \ref{FullScan_yeym}, where
the blue dots satisfy the current bounds.
The upper bound of $y_\mu$ is determined by the data  on 
$\sigma(e^+e^-\rightarrow\mu^+\mu^-)$ (\ref{sigma-mm}), as indicated in Figures \ref{189yeym}, \ref{189ymML} and \ref{189ymuMx}.
Meanwhile, the lower and upper bounds of $y_e$ are dictated by the constraint on $\Delta a_e$ (\ref{delta-ae}), and the LEP data on
$\sigma(e^+e^-\rightarrow e^+e^-)$ (\ref{sigma-ee}) and
$\sigma(e^+e^-\rightarrow \mu^+\mu^-)$ (\ref{sigma-mm}), as shown in Figures
\ref{189yeym}, \ref{189yeML}, and \ref{189yeMx}.
Comparing with the green region in Figure \ref{FCCyeym}, we see that
the viable region in Figure \ref{FullScan_yeym} is significantly expanded (especially in the direction of $y_e$) by varying the masses $m_L$ and $m_{\chi_r}$.
In particular, the allowed ranges for $y_e$ and $y_\mu$ are approximately
[0.8, 2.25] and [0, 0.8], respectively.
In Figure \ref{FullScan_yeym}, the red dots satisfy both current constraints and expected FCC-ee bounds.
After getting results from the FCC-ee, the expected allowed ranges for $y_e$ and $y_\mu$ are about
[1.1, 1.5] and [0, 0.2].
They are significantly narrower than the current allowed ranges of these parameters.

\section{Conclusions}


Vectorlike fermions are symmetric objects in terms of their left and right component behaviors under gauge transformations.
Despite this beauty, their existence is still questionable.
Considering an SM extension with a vectorlike lepton doublet and two complex scalars, we have investigated possible new physics imprints on lepton anomalous magnetic moments, cross sections and forward-backward asymmetries of electron and muon pair production processes at $e^+e^-$ colliders.
Taking into account various current constraints including the measurement of electron and muon magnetic moment, the LEP and the LHC data, as well as the perturbation limit on exotic Yukawa couplings, the viable parameter regions of the model have been identified.
It has been shown that the LEP measurements on cross sections and forward-backward asymmetries of the above two processes play an important role beside the general LEP mass condition on new charged fermions.
The mass hierarchy among $\chi_r$, $\chi_i$, and $L$ turns out to be essential when applying the search result for sleptons pair production at the LHC.
Due to the high precision expected at the FCC-ee, this future collider will be able to exclude a large portion of the parameter space, pinpointing a very narrow viable region to test the model in the future.



\section*{Acknowledgements}
This research is funded by Vietnam 
National Foundation for Science and Technology Development (NAFOSTED)
under grant number 103.01-2023.75.

\bibliographystyle{JHEP} 
\bibliography{references}
\end{document}